\documentclass[12pt]{article}

\usepackage[utf8]{inputenc}
\usepackage{geometry}
\usepackage{graphicx}
\usepackage[outdir=./]{epstopdf}
\usepackage{array}
\usepackage{subfig}
\usepackage{slashed}
\usepackage{amsmath}
\usepackage{amsfonts}
\usepackage{stmaryrd}
\SetSymbolFont{stmry}{bold}{U}{stmry}{m}{n} 
\usepackage{amssymb}
\usepackage[cm]{fullpage}
\usepackage{cite}
\usepackage{epsfig}
\usepackage{comment}
\usepackage{hyperref}
\usepackage{multirow}
\usepackage{cancel}
\usepackage{mathrsfs}

\usepackage{cleveref}
\crefname{section}{Â§}{Â§Â§}
\Crefname{section}{Â§}{Â§Â§}

\numberwithin{equation}{section}

\def\p{\partial}

\def\0{{(0)}}
\def\1{{(1)}}
\def\2{{(2)}}

\def\<{\langle }
\def\>{\rangle }

\newcommand{\bea}{\begin{eqnarray}}

\newcommand{\eea}{\end{eqnarray}}

\newcommand{\be}{\begin{equation}}
\newcommand{\ee}{\end{equation}}
\newcommand{\ba}{\begin{align}}
\newcommand{\ea}{\end{align}}

\makeatletter
  \let\over=\@@over \let\overwithdelims=\@@overwithdelims
  \let\atop=\@@atop \let\atopwithdelims=\@@atopwithdelims
  \let\above=\@@above \let\abovewithdelims=\@@abovewithdelims
\renewcommand\section{\@startsection {section}{1}{\z@}%
                                   {-3.5ex \@plus -1ex \@minus -.2ex}
                                   {2.3ex \@plus.2ex}%
                                   {\normalfont\large\bfseries}}

\renewcommand\subsection{\@startsection{subsection}{2}{\z@}%
                                     {-3.25ex\@plus -1ex \@minus -.2ex}%
                                     {1.5ex \@plus .2ex}%
                                     {\normalfont\bfseries}}
\makeatother

\newcommand{\beq}{\begin{equation}}
\newcommand{\eeq}{\end{equation}}
\newcommand{\beqa}{\begin{eqnarray}}
\newcommand{\eeqa}{\end{eqnarray}}
\newcommand{\beqar}{\begin{eqnarray*}}

\newcommand{\mc}[1]{\mathcal{#1}}

\def\[{\[}
\def\]{\]}

\newcommand{\bd}[1]{\begin{fmffile}{#1}\begin{fmfgraph*}}
\newcommand{\ed}{\end{fmfgraph*}\end{fmffile}}

\newcommand{\mN}{\mathcal{N}}
\newcommand{\N}{\mathcal{N}}
\newcommand{\mb}{\mathbb}

\newcommand{\Nstar}{\mN=2^\star}
\newcommand{\Pa}{\partial}
\newcommand{\Bp}{\bar\partial}

\newcommand{\sigsig}[3]{(\sigma^{#1}\bar\sigma^{#2})_{#3}}
\newcommand{\ssss}[5]{(\sigma^{#1}\bar\sigma^{#2}\sigma^{#3}\bar\sigma^{#4})_{#5}}

\usepackage{slantsc}
\usepackage{booktabs}

\begin{document}

\pagenumbering{Alph} 
\begin{titlepage}

\unitlength = 1mm~
\vskip 2cm
\begin{center}

{\Large{{\sc $\Omega$ } {\it versus } \textsc{Graviphoton}}}

\vspace{0.8cm}
Ahmad Zein Assi\,\footnote{\tt zeinassi@cern.ch}

\vspace{1cm}

{\it Albert Einstein Center for Fundamental Physics\\
Institute for Theoretical Physics, University of Bern\\
\vspace{0.4cm}
Sidlerstrasse 5, CH-3012 Bern, Switzerland 

}

\vspace{0.8cm}

\begin{abstract}
In this work, I study the deformation of the topological string by $\bar\Omega$, the complex conjugate of the $\Omega$-deformation. Namely, I identify $\bar\Omega$ in terms of a physical state in the string spectrum and verify that the deformed Yang-Mills and ADHM actions are reproduced. This completes the study initiated in \cite{BarDec} where we show that $\bar\Omega$ decouples from the one-loop topological amplitudes in heterotic string theory. Similarly to the $\Nstar$ deformation, I show that the quadratic terms in the effective action play a crucial role in obtaining the correct realisation of the full $\Omega$-deformation. Finally, I comment on the differences between the graviphoton and the $\Omega$-deformation in general and discuss possible $\bar\Omega$ remnants at the boundary of the string moduli space.

\end{abstract}

\vspace{1.0cm}

\end{center}

\end{titlepage}

\pagenumbering{arabic} 

\pagestyle{empty}
\pagestyle{plain}

\def\vx{{\vec x}}
\def\p{\partial}
\def\po{$\cal P_O$}

\pagenumbering{arabic}

\tableofcontents

\section{Introduction}

Little attention has been devoted to the study of the $\Omega$-background as a non-holomorphic deformation at the string level. Indeed, from the gauge theory point of view, the $\Omega$-deformation is a background that twists space-time in a particular way allowing for a path-integral derivation of the partition function through localisation \cite{Losev:1997wp,Nekrasov:2002qd,Nekrasov:2003rj}. Viewed as a background, it is non-holomorphic in the sense that it has a holomorphic as well as an anti-holomorphic part denoted $\bar\Omega$. However, the latter decouples from physical observables and this is clear from the localisation perspective where $\bar\Omega$ is a Q-exact deformation of the effective action.

From the string theory point of view, there is a natural \emph{topological} limit that one can impose on the $\Omega$-background. If one denotes $\epsilon_{1,2}$ the two parameters of the latter, then, for $\epsilon_1+\epsilon_2=0$ the $\Omega$-deformed gauge theory partition function is the field theory limit of the topological string partition function $F_g$ \cite{Antoniadis:1995zn} which computes a class of higher derivative gravitational couplings in the effective action \cite{Antoniadis:1993ze}. The connection between string amplitudes and supersymmetric gauge theories has been further extended beyond topological limit \cite{AFHNZ,Antoniadis:2013mna,Antoniadis:2015spa}. In all these studies, the anti-holomorphic part $\bar\Omega$ is implicitly set to zero. This is understandable from the gauge theory side because of the decoupling of $\bar\Omega$. In string theory, however, it is not clear, \emph{a priori}, that the same property remains true. A particular instance where the breakdown, in string theory, of gauge theory properties is the moduli dependence of topological amplitudes. Indeed, as shown in \cite{Bershadsky:1993cx}, the topological amplitudes $F_g$ have a non-holomorphic dependence on the moduli of the string compactification stemming from the boundary of the moduli space only while, in gauge theory, $F_g$ is purely holomorphic.

The purpose of the present paper is to initiate a similar analysis for the $\Omega$-deformation in the topological limit. The goal is to study the possible dependence in string theory on $\bar\Omega$ by identifying it in the string theory spectrum. In \cite{BarDec}, by making a particular ansatz for $\bar\Omega$, we showed that $\bar\Omega$ decouples from the topological amplitudes $F_g$ perturbatively. Here, I prove that this ansatz is correct by showing that it leads to the correct effective actions in field theory. This is done by coupling the open string degrees of freedom of a Dp-D(p+4) system in type I string theory to the closed string background using similar techniques as in \cite{Billo:2006jm,Ito:2010vx,Antoniadis:2013mna}.

The paper is organised as follows. In Section \ref{DefOmegaBar}, I review the ansatz of \cite{BarDec} in the context of type I string theory compactified on a $T^2\times K3$ orientifold and then, in Section \ref{TypeI}, I derive the perturbative (Yang-Mills) and non-perturbative (ADHM) effective actions in the presence of the string theory $\Omega,\bar\Omega$-background. Finally, in Section \ref{Het}, I briefly review the results of \cite{BarDec} for completeness. To keep the discussion clear, several technical aspects and useful results are presented in two appendices.


\section{\texorpdfstring{The stringy $\bar\Omega$}{The Stringy OmegaBar}}\label{DefOmegaBar}

In \cite{BarDec}, starting from heterotic string theory compactified on $T^2\times K3$, we postulated that $\bar\Omega$ can be identified as a constant background for a self-dual field strength $F_T$ which is the vector partner of the K\"ahler modulus $T$ of $T^2$. Here, I repeat the same analysis in the context of the dual type I string theory.

Recall that, in ten dimensions, type I and heterotic string theories are S-dual to each other. However, in four dimensions, there are regions in the moduli space where both theories are weakly coupled \cite{Antoniadis:1997gu}. Hence, I focus on those particular regions and consider a type I theory compactified on $T^2\times K3$. Here, I realise $K3$ as a $T^4/\mb Z_2$ orientifold which admits both D9- and D5-branes. Tadpole cancellation restricts the number of such D-branes. However, I keep their number generic as one could consider a non-compact $K3$ as well. In table \ref{tb:HetTypeI}, I summarise the mapping between the universal scalar fields of the vector multiplet moduli space.

\begin{table}[ht]
 \begin{center}
  \begin{tabular}{cc}
  \toprule
  Heterotic &Type I\\\midrule
  S&S\\
  T&S'\\
  U&U\\\bottomrule
  \end{tabular}
 \end{center}
\caption{Mapping of the universal fields under Heterotic/Type I duality in four dimensions.}
\label{tb:HetTypeI}
\end{table}

Notice that the axion-dilatons are mapped to one another \cite{Antoniadis:1997gu,Antoniadis:1997nz}. Furthermore, the complex structure modulus U of $T^2$ is unchanged and the K\"ahler modulus T in Heterotic is mapped to another dilaton-like field denoted $S'$. Indeed, the presence of two dilatons in Type I is not surprising since the theory contains D9- and D5-branes whose coupling constants are given by the imaginary parts of two different fields, namely $S$ and $S'$ respectively.

In order to identify the $\bar\Omega$ deformation in terms of the type I physical string spectrum, I focus on the universal vector multiplet sector, \emph{i.e.} the so-called STU-model. Namely, the three vector multiplet moduli are S, S' and U. To keep the discussion clear, I do not consider Wilson lines. In addition to the three vector fields associated to each scalar, there is another vector field stemming from the $\N=2$ gravity multiplet, the graviphoton, so that there are four gauge fields denoted $F_I$ with $I=G,S,S',U$.

Since the question under consideration is the realisation of the $\Omega$-deformation, recall that one can construct it geometrically as a non-trivial $T^2$ fibration over space-time, in such a way that, when one goes around the cycles of $T^2$, the space-time fields are rotated with an arbitrary angle. This geometric picture of a reduction from six dimensions on the metric with line element
\begin{equation}\label{OmegaMetric}
\textrm{d}s^2=g_{\mu\nu}\left(\textrm{d}X^\mu+{\Omega^\mu}_\rho X^\rho \textrm{d}Z+{\bar\Omega^\mu}_\rho X^\rho \textrm{d}\bar Z\right)\left(\textrm{d}X^\nu+{\Omega^\nu}_\rho X^\rho \textrm{d}Z+{\bar\Omega^\nu}_\rho X^\rho \textrm{d}\bar Z\right)+\textrm{d}s^2_{T^2}
\end{equation}
is very helpful. Here, $\mu,\nu\in[\![0,3]\!]$ are space-time indices, $Z$ is the complexified $T^2$ coordinate and $\Omega,\bar\Omega$ are the space-time rotation matrices depending on two complex parameters $\epsilon_{1,2}$. Hence, the Lorentz group $SU(2)_L\times SU(2)_R$ is explicitly broken. For practical purposes, redefine them as
\begin{equation}
 \epsilon_{\pm}=\frac{\epsilon_1\pm\epsilon_2}{2}.
\end{equation}
For the present matter, I work in the topological limit $\epsilon_+=0$. From the seminal work \cite{Antoniadis:1993ze}, one can show that the holomorphic part of the $\Omega$-deformation can be described as a constant background for the self-dual part of the graviphoton field strength. The self-duality condition means that $\epsilon_-\equiv\epsilon$ is sensitive to $SU(2)_L$ only. In this convention, for $\epsilon_+=0$, $SU(2)_R$ remains unbroken. Consequently, in order to describe $\bar\Omega$ in type I string theory, one should turn on a constant background for the self-dual field strength of one of the following fields.

\begin{enumerate}
 \item ${F_{S'}}^{(-)}$: $V_T=\epsilon_\mu\partial X^\mu\bar\partial\bar Z$+\textrm{RR}\,.
 \item ${F_U}^{(-)}$: $V_U=\epsilon_\mu\partial X^\mu\bar\partial Z$+\textrm{RR}\,.
 \item ${F_S}^{(-)}$: $V_S=\epsilon_\mu\partial\bar Z\bar\partial X^\mu$+\textrm{RR}\,.
\end{enumerate}

The superscript $(-)$ stresses the fact that the field strengths are self-dual. Notice that I have only written the bosonic part of the vertex operators which should be completed by the fermionic and the Ramond-Ramond parts\footnote{The full vertex operators are written in Section \ref{TypeI}.}. As opposed to heterotic, the latter arises typically in orientifold compactifications since the supercharges are combinations of left- and right-moving ones. In addition, I have implicitly assumed that the polarisation $\epsilon_\mu$ is chosen in such a way to satisfy the self-duality constraint. Recall that the graviphoton vertex operator is
\begin{equation}
 V_G=\epsilon_\mu\partial Z\bar\partial X^\mu+\textrm{RR}\,.
\end{equation}
Consequently, one is led to choosing ${F_{S'}}^{(-)}$ and ${F_S}^{(-)}$ as natural candidates for $\bar\Omega$. Here, `natural' means that $\bar\Omega$ is understood, in some sense, as the complex conjugate of $\Omega$.

In order to rigorously make the selection, I first consider the effect of each of these states on the gauge theory degrees of freedom. More precisely, following the ideas pioneered in \cite{Green:2000ke} and further exploited in \cite{Billo:2006jm,Ito:2010vx,Antoniadis:2013mna}, I analyse the possible couplings of the field-strengths with the gauge theory degrees of freedom (including instantons) realised as the massless excitations of the open strings in Dp-D(p+4) branes systems. This is naturally realised in the dual type I string theory \cite{Antoniadis:2013mna} studied in the following section. 


\section{\texorpdfstring{$\bar\Omega$-deformed effective actions}{OmegaBar-deformed effective actions}}\label{TypeI}

\subsection{Vertex Operators}

In the present section, I use the D9-branes to realise the gauge theory. By T-duality, this is equivalent to the D3-branes realisation \cite{Green:2000ke}, see also \cite{These} for a review and complete analysis. In this setup, gauge theory instantons are realised in terms of D5-branes called D5-instantons wrapping $T^2\times K3$. This configuration is summarised in the following table.

\begin{center}
\parbox{14.5cm}{\begin{center}\begin{tabular}{|c||c||c|c|c|c||c|c||c|c|c|c|}\hline
&&&&&&&&&&&\\[-11pt]
\bf{brane} & \textbf{num.} & $X^0$ & $X^1$ & $X^2$ & $X^3$ & $X^4$ & $X^5$ & $X^6$ & $X^7$ & $X^8$ & $X^9$  \\\hline\hline
D9 & $N$ & $\bullet$ & $\bullet$ & $\bullet$ & $\bullet$ & $\bullet$ & $\bullet$ & $\bullet$ & $\bullet$ & $\bullet$ & $\bullet$\\\hline
D5 & $k$ &  &  &  &  & $\bullet$ & $\bullet$ & $\bullet$ & $\bullet$ & $\bullet$ & $\bullet$\\\hline
\end{tabular}\end{center}
${}$\\[-48pt]
\begin{align}
{}\hspace{2.9cm}\underbrace{\hspace{3.8cm}}_{\text{space-time}\,\sim\,\mathbb{R}^4}\underbrace{\hspace{1.9cm}}_{T^2}\underbrace{\hspace{3.75cm}}_{K3\,\sim\, T^4/\mathbb{Z}_2}\nonumber
\end{align}}
\end{center}
In this picture, the massless excitations of the open strings stretched between the D9-branes are the gauge theory degrees of freedom whereas the massless excitations of the open strings with at least one endpoint on a D5-instanton correspond to the ADHM moduli. Hence, besides the D9-branes excitations that I refer to as the 9-9 sector, there are two classes of open string excitations in the instanton sector, depending on whether the location of one endpoint of the open string only lies on a D5-instanton (9-5, 5-9 or mixed sector) or both (5-5 or unmixed sector). The ADHM moduli are non dynamical fields due to the Dirichlet boundary conditions in all transverse directions. These fields are summarised in Table \ref{Tab:Fields}.

\begin{table}[ht]
\begin{center}
\begin{tabular}{|c||c||c|c|c|}\hline
&&&&\\[-11pt]
\bf{sector} & \bf{field} & $SO(4)_{ST}$ & $c_{T^2}$ & $SU(2)_+\times SU(2)_-$ \\\hline\hline
&&&&\\[-11pt]
9-9 & $A^\mu$ & $(\mathbf{1/2},\mathbf{1/2})$ & $0$ & $(\mathbf{1},\mathbf{1})$ \\
&&&&\\[-11pt]\hline
&&&&\\[-11pt]
 & $\Lambda^{\alpha A}$ & $(\mathbf{1/2},\mathbf{0})$ & $1/2$ & $(\mathbf{2},\mathbf{1})$ \\
 &&&&\\[-11pt]\hline
 &&&&\\[-11pt]
 & $\Lambda_{\dot{\alpha} A}$ & $(\mathbf{0},\mathbf{1/2})$ & $-1/2$ & $(\mathbf{2},\mathbf{1})$ \\
 &&&&\\[-11pt]\hline
 &&&&\\[-11pt]
 & $\phi$ & $(\mathbf{0},\mathbf{0})$ & $-1$ & $(\mathbf{1},\mathbf{1})$ \\ &&&&\\[-11pt]\hline\hline
 &&&&\\[-11pt]
 5-5 & $a^\mu$ & $(\mathbf{1/2},\mathbf{1/2})$ & $0$ & $(\mathbf{1},\mathbf{1})$  \\
&&&&\\[-11pt]\hline
&&&&\\[-11pt]
 & $\chi$ & $(\mathbf{0},\mathbf{0})$ & $-1$ & $(\mathbf{1},\mathbf{1})$ \\
&&&&\\[-11pt]\hline
&&&&\\[-11pt]
& $M^{\alpha A}$ & $(\mathbf{1/2},\mathbf{0})$ & $1/2$ & $(\mathbf{2},\mathbf{1})$ \\
&&&&\\[-11pt]\hline
&&&&\\[-11pt]
& $\lambda_{\dot{\alpha} A}$ & $(\mathbf{0},\mathbf{1/2})$ & $-1/2$ & $(\mathbf{2},\mathbf{1})$ \\
&&&&\\[-11pt]\hline\hline
&&&&\\[-11pt]
5-9 & $\omega_{\dot{\alpha} }$ & $(\mathbf{0},\mathbf{1/2})$ & $0$ & $(\mathbf{1},\mathbf{1})$ \\
&&&&\\[-11pt]\hline
&&&&\\[-11pt]
 & $\mu^A$ & $(\mathbf{0},\mathbf{0})$ & $1/2$ & $(\mathbf{2},\mathbf{1})$ \\[2pt]\hline
\end{tabular}
\end{center}
\caption{Summary of the massless spectrum of the D5/D9 system and the
decomposition of the scalar fields. I display the transformation properties under the groups $SO(4)_{ST}\times SU(2)_+\times SU(2)_-$, while $c_{T^2}$ is the charge under $SO(2)_{T^2}$.}
\label{Tab:Fields}
\end{table}

In the 9-9 sector, the massless excitations consist of a number of $\mathcal{N}=2$ vector multiplets, each of which containing a vector field $A^\mu$, a complex scalar $\phi$ as well as four gaugini $(\Lambda^{\alpha A}, \Lambda_{\dot\alpha A})$ that are in the $(\mathbf{2},\mathbf{1})$ representation of $SU(2)_{+}\times SU(2)_-$. The bosonic degrees of freedom stem from the NS sector, while the fermionic ones from the R sector. These fields, taken separately, realise a Yang-Mills theory in four dimensions. Their vertex operators are
\begin{align}
 V_{A}(z)&= \frac{A_{\mu}(p)}{\sqrt{2}}\psi^{\mu}(z)\,e^{ip\cdotp X(z)}\,e^{-\varphi(z)}\,,&V_{\phi}(z)&= \frac{\phi_a(p)}{\sqrt{2}}\psi^a(z)\,e^{ip\cdotp X(z)}\,e^{-\varphi(z)}\,,\label{AVop}\\
 V_{\Lambda}(z)&= \Lambda^{\alpha A}S_{\alpha}(z)S_{A}(z)\,e^{ip\cdotp X(z)}\,e^{-\tfrac{1}{2}\varphi(z)}\,,&V_{\bar\Lambda}(z)&= \Lambda_{\dot\alpha A}S^{\dot\alpha}(z)S^{A}(z)\,e^{ip\cdotp X(z)}\,e^{-\tfrac{1}{2}\varphi(z)}\label{LambdaVop}\,.
\end{align}
In the 5-5 and NS sector, there are ten bosonic moduli that can be written as a real vector $a^\mu$ and six scalars $\chi^I$. From the perspective of the gauge theory living on the world-volume of the D9-branes, $a^\mu$ parametrises the position of gauge theory instantons. In the Ramond sector, there are sixteen fermionic moduli $M^{\alpha A}$, $\lambda_{\dot\alpha A}$. The vertex operators of these states are
\begin{align}
 V_{a}(z)&= g_6\,a_{\mu}\psi^{\mu}(z)e^{-\varphi(z)}\,,&V_{\chi}(z)&=
\frac{\chi_I}{\sqrt{2}}\psi^I(z)e^{-\varphi(z)}\,,\label{aVop}\\
 V_{M}(z)&= \frac{g_6}{\sqrt{2}}\,M^{\alpha
A}S_{\alpha}(z)S_{A}(z)e^{-\tfrac{1}{2}\varphi(z)}\,,&V_{\lambda}(z)&=
\lambda_{\dot\alpha
A}\,S^{\dot\alpha}(z)S^{A}(z)\,e^{-\tfrac{1}{2}\varphi(z)}\,.\label{MlambdaVop}
\end{align}
Here, $g_5$ is the D5-brane coupling constant. For a Dp-brane, it is given by
\begin{equation}
 g_{p+1}^2=4\pi(2\pi\sqrt{\alpha'})^{p-3}\,g_s\,.
\end{equation}
Finally, let's consider the mixed moduli. From the NS sector, the fermionic coordinates give rise to two Weyl spinors $(\omega_{\dot\alpha},\bar\omega_{\dot\alpha})$ of $SO(4)$. They have the same chirality due to the specific choice of boundary conditions of the D5-branes. In addition, they describe the size of the instanton. In the R sector, one gets two Weyl fermions $(\mu^A, \bar\mu^A)$ transforming in the fundamental representation of $SO(6)$.
The vertex operators for the mixed sectors contain the \emph{twist operators} that change the coordinates boundary conditions from Dirichlet to Neumann and vice versa. These are bosonic fields $\Delta,\bar\Delta$ of conformal dimension 1/4. The vertex operators are
\begin{align}
 V_{\omega}(z)=&
\frac{g_1}{\sqrt{2}}\,\omega_{\dot\alpha}\Delta(z)S^{\dot\alpha}(z)e^{
-\varphi(z)}\,,& V_{\bar\omega}(z)=&
\frac{g_1}{\sqrt{2}}\,\bar\omega_{\dot\alpha}\bar\Delta(z)S^{\dot\alpha}(z)e^{
-\varphi(z)}\,\label{OmegaVop}\\
 V_{\mu}(z)=&
\frac{g_1}{\sqrt{2}}\,\mu^A\Delta(z)S_{A}(z)e^{-\frac{1}{2}\varphi(z)}\,,&V_{
\bar\mu}(z)=&
\frac{g_1}{\sqrt{2}}\,\bar\mu^{A}\bar\Delta(z)S_{A}(z)e^{-\frac{1}{2}\varphi(z)}
\,.\label{muVop}
\end{align}

This summarises the complete description of the gauge theory states in terms of CFT vertex operators. In order to study the $\bar\Omega$-deformation, one must include the closed string operators describing it. As mentioned above, the preserved supercharges are a combination of the left- and right-moving supercharges due to the orientifold action. This implies that the vertex operators for the vector fields have, generically, an NS-NS and a R-R part each. They can be derived explicitly by spectral flow from the universal scalars' vertex operators. In \cite{AFHNZ}, the self-dual graviphoton and U-vector field operators were shown to take the form
\begin{align}
	V^{\pm}(y,\bar y)=\frac{1}{8\pi\,\sqrt{2}}F_{\mu\nu}^{\pm} \bigg[\,\psi^{\mu}\psi^{\nu}(y)e^{-\varphi(\bar y)}{\Psi}(\bar y)\,+e^{-\varphi(y)}\Psi(y)\psi^{\mu}\psi^{\nu}(\bar y)&\nonumber\\
\pm\frac{i}{2}\, e^{-\tfrac{1}{2}(\varphi(y)+\varphi(\bar y))} S_{\alpha}(y) (\sigma^{\mu\nu})^{\alpha\beta} 
S_{\beta}(\bar y)\,\epsilon^{AB}\,S_{A}(y)S_{B}(\bar y)\bigg]&\,,\label{GraviphotonVertex}
\end{align}
with $V_{G}=V^-$, $V_{U}=V^+$, $F^{-}={F_G}^{(-)}$ and $F^{+}={F_U}^{(-)}$. Also note that $\Psi$ is the worldsheet fermion in the $T^2$ direction. It is important to notice that the difference between the two states is merely the sign between the NS-NS and R-R parts. Similarly, the vertex operators for the candidate states, \emph{i.e.} ${F_S}^{(-)}$ and ${F_{S'}}^{(-)}$ are found to be
\begin{align}
	\bar V^{\pm}(y,\bar y)=\frac{1}{8\pi\,\sqrt{2}}\bar F_{\mu\nu}^{\pm} \bigg[\,\psi^{\mu}\psi^{\nu}(y)e^{-\varphi(\bar y)}{\bar\Psi}(\bar y)\,+e^{-\varphi(y)}\bar\Psi(y)\psi^{\mu}\psi^{\nu}(\bar y)&\nonumber\\
\pm\frac{i}{2}\, e^{-\tfrac{1}{2}(\varphi(y)+\varphi(\bar y))} S_{\alpha}(y) (\sigma^{\mu\nu})^{\alpha\beta} 
S_{\beta}(\bar y)\,\epsilon^{\hat A\hat B}\,S_{\hat A}(y)S_{\hat B}(\bar y)\bigg]&\,,\label{SVertex}
\end{align}
where $V_{S}=\bar V^+$, $V_{S'}=\bar V^-$, $\bar F^{+}={F_S}^{(-)}$ and $\bar F^{-}={F_{S'}}^{(-)}$.

\subsection{Deformed Yang-Mills action}

I now calculate the deformations to the Yang-Mills effective action due to the candidates for the $\bar\Omega$-deformation. In order to achieve this, I calculate all possible tree-level couplings between the self-dual S- and S'-vectors to leading order in $\alpha'$. In the type I theory under study, these are simply disc diagrams whose boundary lies entirely on a D9-brane. In addition, I insert at least one self-dual S- or S'-vector vertex operator in the bulk of the disc diagram (and possibly some graviphotons) while including an arbitrary number of open string insertions at the boundary from the 9-9 sector. In fact, the number of such insertions is highly restricted by the fact that I am only interested in the gauge theory action.

\begin{figure}[h!t]
\begin{center}
\includegraphics[width=0.8\textwidth]{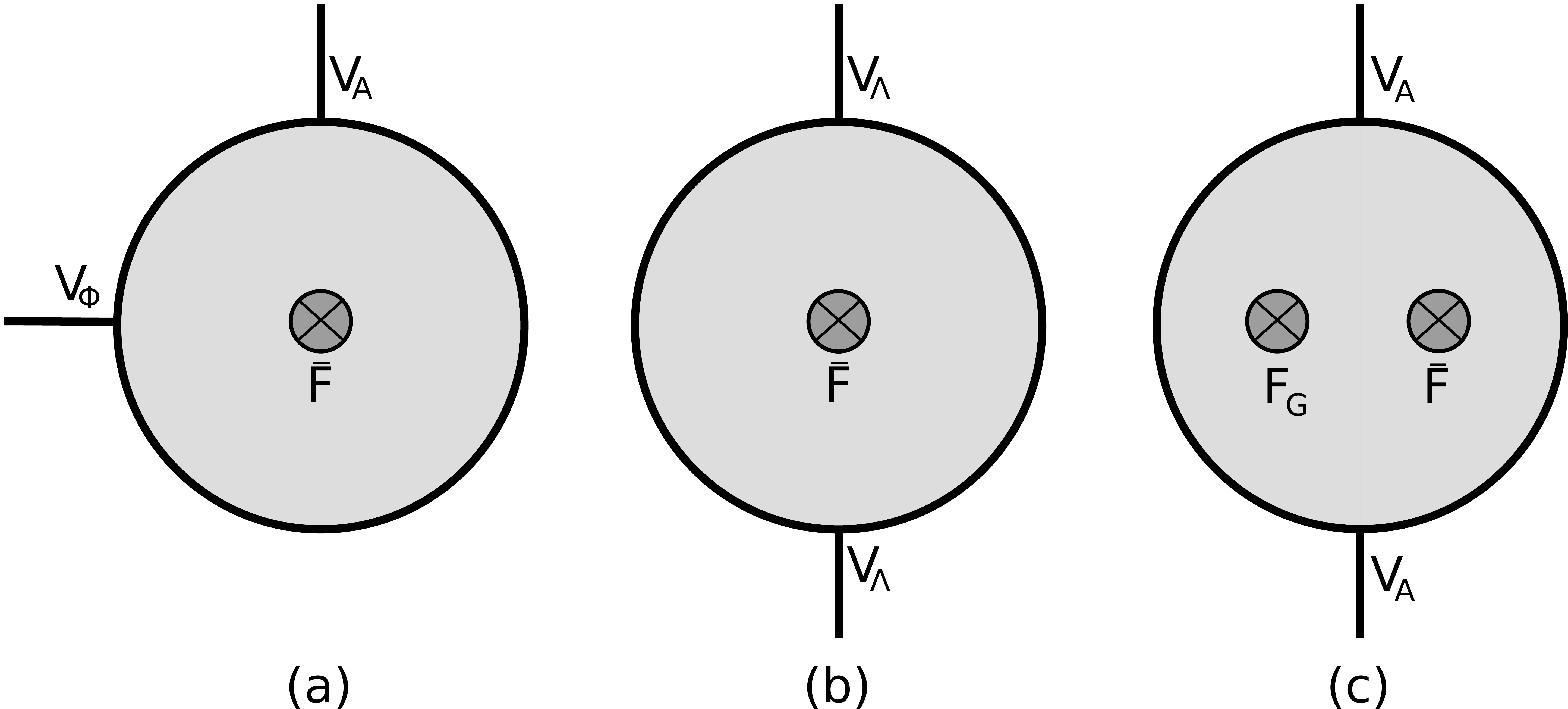}
\end{center}
\caption{Disc diagrams in the 9-9 sector. Diagram (a) involves two bosonic boundary insertions and one bulk insertion of $\bar F$, diagram (b) two fermionic insertions and one $\bar F$, whereas diagram (c) has two bosonic boundary insertions and two bulk insertions of a graviphoton and an $\bar F$.} 
\label{fig:YMdiags}
\end{figure}

From a practical point of view, to avoid repeating the calculations twice, I split the vertex operator for the $\bar\Omega$-deformation into an NS-NS part $V_1$ and a R-R one $V_2$. It turns out that only few disc amplitudes are potentially non-vanishing in the field theory limit and some of those are summarised in Fig. \ref{fig:YMdiags}.

I start by evaluating the amplitude with the gaugini. Notice that, due to the specific self-duality of the closed string vertex, the only possible non-trivial couplings are with $\Lambda^{\alpha A}$. Furthermore, had one used the graviphoton vertex instead, the amplitude would have vanished trivially by the non-conservation of the $SO(2)_{T^2}$-charge. Using the doubling trick, I convert the disc into the full plane with a $\mb Z_2$ involution. I then split the closed string vertices into their left- and right-moving parts which can be considered as independent. As such, the tree-point disc amplitude becomes effectively a four-point amplitude. In addition, I soak up $c$-ghost zero-modes on the sphere by attaching $c$ to three dimension-one operators in such a way that the resulting vertex operator is BRST closed. In order to avoid subtleties due to the use of unintegrated vertex operators in different ghost pictures as discussed, for instance, in \cite{These}, I don't attach the $c$-ghost to operators in the zero-picture. The contribution from the NS-NS part of $\bar F$ is
\begin{align}\label{LambdaLambdaFsetup}
 \left\langle\left\langle c V_{\Lambda}(z_1)c V_{\Lambda}(z_2)c V^{(-1,0)}_{\bar F}(z_3,z_4)\right\rangle\right\rangle&=\frac{1}{8\pi\,\sqrt{2}}\Lambda^{\alpha A}\Lambda^{\beta B}\bar F_{\mu\nu}\Big\langle\Big\langle c(z_1)S_{\alpha}(z_1)S_{A}(z_1)\,e^{-\tfrac{1}{2}\varphi(z_1)}\nonumber\\
&\times c(z_2)S_{\beta}(z_2)S_{B}(z_2)\,e^{-\tfrac{1}{2}\varphi(z_2)}\nonumber\\
&\times c(z_3)e^{-\varphi(z_3)}\bar\Psi(z_3)\psi^{\mu}\psi^{\nu}(z_4)\Big\rangle\Big\rangle+(\textrm{Left}\leftrightarrow\textrm{Right})\,.
\end{align}
Notice that I have denoted the left and right parts of the NS-NS vertex by two different labels. The double brackets notation means that one should integrate over $n-3$ free positions with $n$ being the total number of inserted points. In \eqref{LambdaLambdaFsetup}, one must integrate over $z_4$ only. Performing the contractions between the various operators using the standard results summarised in Appendix \ref{appendix:spinors}, I find
\begin{align}
 \left\langle\left\langle c V_{\Lambda}(z_1)c V_{\Lambda}(z_2)c V^{(-1,0)}_{\bar F}(z_3,z_4)\right\rangle\right\rangle&=-\frac{i}{16\pi\,\sqrt{2}}\epsilon_{AB}(\sigma^{\mu\nu})_{\alpha\beta}\Lambda^{\alpha A}\Lambda^{\beta B}\bar F_{\mu\nu}\int\textrm{d}z^4\frac{z_{12}}{z_{14}z_{24}}\,.
\end{align}
I have used the notation $z_{ij}\equiv z_i-z_j$. The integral over $z^4$ is straightforward. I identify $\bar\epsilon$ using a particular normalisation for $\bar F_{\mu\nu}$:
\begin{equation}
\bar F_{\mu\nu}\equiv \eta^3_{\mu\nu}\frac{\bar\epsilon}{2}\,, 
\end{equation}
with $\eta^{c}_{\mu\nu}$ being the 't Hooft symbols. Thus, I obtain
\begin{align}
 \left\langle\left\langle c V_{\Lambda}(z_1)c V_{\Lambda}(z_2)c V^{(-1,0)}_{\bar F}(z_3,z_4)\right\rangle\right\rangle&=-i\frac{\bar\epsilon}{\sqrt{2}}\Lambda^{\alpha A}{\Lambda^\beta}_A(\sigma^3)_{\alpha\beta}\,.
\end{align}
I now turn to the contribution of the R-R piece of the closed string operator which is
\begin{align}
 \left\langle\left\langle c V_{\Lambda}(z_1)c V_{\Lambda}(z_2)c V^{(-\frac{1}{2},-\frac{1}{2})}_{\bar F}(z_3,z_4)\right\rangle\right\rangle&=\frac{i}{16\pi\,\sqrt{2}}\Lambda^{\alpha A}\Lambda^{\beta B}\bar F_{\mu\nu}(\sigma^{\mu\nu})^{\gamma\delta}\epsilon^{\hat C\hat D}\Big\langle\Big\langle c(z_1)S_{\alpha}(z_1)S_{A}(z_1)e^{-\tfrac{1}{2}\varphi(z_1)}\nonumber\\
&\times c(z_2)S_{\beta}(z_2)S_{B}(z_2)\,e^{-\tfrac{1}{2}\varphi(z_2)}c(z_3)e^{-\tfrac{1}{2}\varphi(z_3)} S_\lambda(z_3)S_{\hat C}(z_3)\nonumber\\
&\times e^{-\tfrac{1}{2}\varphi(z_3)}S_{\delta}(z_4)S_{\hat D}(z_4)\Big\rangle\Big\rangle+(\textrm{Left}\leftrightarrow\textrm{Right})\,.
\end{align}
After making all possible contractions, the result is
\begin{align}
 \left\langle\left\langle c V_{\Lambda}(z_1)c V_{\Lambda}(z_2)c V^{(-\frac{1}{2},-\frac{1}{2})}_{\bar F}(z_3,z_4)\right\rangle\right\rangle&=-\frac{i}{16\pi\,\sqrt{2}}\Lambda^{\alpha A}\Lambda^{\beta B}\bar F_{\mu\nu}(\sigma^{\mu\nu})^{\gamma\delta}\epsilon^{\hat C\hat D}\epsilon_{\alpha\delta}\epsilon_{\beta\gamma}\epsilon_{AB}\epsilon_{\hat C\hat D}\int\textrm{d}z^4\frac{z_{12}}{z_{14}z_{24}}\,.
\end{align}
Therefore, one gets exactly the same result as for the NS-NS contribution:
\begin{align}
 \left\langle\left\langle c V_{\Lambda}(z_1)c V_{\Lambda}(z_2)c V^{(-\frac{1}{2},-\frac{1}{2})}_{\bar F}(z_3,z_4)\right\rangle\right\rangle&=-i\frac{\bar\epsilon}{\sqrt{2}}\Lambda^{\alpha A}{\Lambda^\beta}_A(\sigma^3)_{\alpha\beta}\,.
\end{align}
This is to be expected based on supersymmetry arguments. One now puts both results together and notices a clear difference between the $S$- and the $S'$-vectors. Namely, for the $F_S$ one has
\begin{equation}
 \Big\langle V_{\Lambda}V_{\Lambda}V_S\Big\rangle=0
\end{equation}
while for $F_{S'}$
\begin{equation}
 \Big\langle V_{\Lambda}V_{\Lambda}V_{S'}\Big\rangle=-i\sqrt{2}\,\bar\epsilon\,\Lambda^{\alpha A}{\Lambda^\beta}_A(\sigma^3)_{\alpha\beta}\,.
\end{equation}
In fact, to the leading order I am concerned by, there are no further couplings between the closed string vertices and the fermionic fields.

This is already a clear indication that $\bar\Omega$ is realised by the $S'$-vector as I now confirm from the following calculations. Note that this type of cancellations was already observed in a similar setup in \cite{Antoniadis:2013mna} where the problem of the refinement is addressed.

Now consider the possible couplings with the bosonic open string degrees of freedom. The first non-trivial coupling corresponds to diagram (a) in Figure \ref{fig:YMdiags}. As before, I split the closed string vertex into its NS-NS and R-R parts. The setup for the vertex operators is as follows.
\begin{align}
 V_{\phi}(z_1)&=\frac{\phi}{\sqrt{2}}c\,e^{-\varphi}\psi\,e^{2ip_1\cdot X}(z_1)\,,\nonumber\\
 V_{A}(z_2)&=A_\mu\left(\partial X^\mu-2i(p_2\cdot\psi)\psi^\mu\right)e^{2ip_2\cdot X}(z_2)\,,\nonumber\\
 V_{\bar F}(z_3)&=c\,e^{-\tfrac{\varphi}{2}}S_\alpha S_{\hat A}e^{i\bar P\cdot X}(z_3)\,,\nonumber\\
 V_{\bar F}(z_4)&=\frac{i}{16\pi\sqrt{2}}\epsilon^{\hat A\hat B}(\sigma^{\lambda\nu})^{\alpha\beta}\bar F_{\lambda\nu}e^{-\tfrac{\varphi}{2}}S_\beta S_{\hat B}e^{i\bar P\cdot X}(z_4)\,.
\end{align}
Hence, I choose to integrate over $z_2$ only. Notice that only the fermion bilinear term of the gauge field can lead to a non-vanishing amplitude. In addition, one can already set the momenta of the exponentials to zero since this amplitude turns out not to stem from contact terms. The correlator is thus
\begin{align}
 \left\langle V_{\phi}(z_1)V_{A}(z_2)V^{(-\tfrac{1}{2},-\tfrac{1}{2})}(z_3,z_4)\right\rangle=&\frac{1}{16\pi}\epsilon^{\hat A\hat B}\phi\,p_{2\rho}A_\mu\bar F^{\alpha\beta}\frac{z_{13}z_{14}z_{24}}{z_{13}^{1/2}z_{14}^{1/2}z_{34}^{1/4}}\nonumber\\
 &\times\left\langle\psi(z_1)\psi^{\rho}\psi^{\mu}(z_2)S_{\alpha}S_{\hat A}(z_3)S_{\beta}S_{\hat B}(z_4)\right\rangle\,,
\end{align}
where I have introduced the notation $F^{\alpha\beta}\equiv(\sigma^{\mu\nu})^{\alpha\beta}F_{\mu\nu}$. Using the results of Appendix \ref{appendix:spinors}, one can calculate the CFT correlator and find
\begin{align}
 \left\langle V_{\phi}(z_1)V_{A}(z_2)V^{(-\tfrac{1}{2},-\tfrac{1}{2})}(z_3,z_4)\right\rangle&=-\frac{i}{16\pi}\phi\,p_{2\rho}A_\mu\bar F^{\alpha\beta}(\sigma^{\rho\mu})_{\alpha\beta}\frac{z_{34}}{z_{23}z_{24}}\,.
\end{align}
Integrating over the disc, one obtains
\begin{align}
 \left\langle\left\langle V_{\phi}(z_1)V_{A}(z_2)V^{(-\tfrac{1}{2},-\tfrac{1}{2})}(z_3,z_4)\right\rangle\right\rangle&=\phi F_{\mu\nu}\bar F^{\mu\nu}\,.
\end{align}
Including the NS-NS part, the total contribution to the action becomes
\begin{align}
  \left\langle V_{\phi}V_{A}V_{S'}\right\rangle&=\bar\epsilon\,\phi F_{\mu\nu}\eta^{3\mu\nu}\,.
\end{align}

Finally, I analyse the possible quadratic terms in the $\Omega$-deformation. I first consider possible mass terms for the gauge fields, namely I calculate the coupling between two gauge fields, one graviphoton and one $S'$-vector,
\begin{equation}
 \left\langle V_{A}(z_1)V_{A}(z_2)V_G(z_3,z_4)V_{S'}(z_5,z_6)\right\rangle\,,
\end{equation}
and choose for convenience to fix the positions $z_1$, $z_2$ and $z_4$ to $-\infty$, $1$ and $-ix$ with $x$ being real (so that $z_3=i x$). This is done by attaching a $c$-ghost to the corresponding vertex operators. In order to keep the calculation simple, I take the gauge fields' vertex operators in the $0$-ghost picture. However, by BRST invariance, this requires including an additional contribution to the vertex \eqref{AVop} as discussed in \cite{Antoniadis:2013mna,These}. Hence, the vertex operators are explicitly given by
\begin{align}
 V_{A}(z_1)&=A_\mu\left[c(\partial X^\mu+2i\,p_{1,\nu}\psi^\mu\psi^\nu)-\gamma\psi^\mu\right]\,e^{2ip_1\cdot X}(z_1)\,,\\
 V_{A}(z_2)&=A_\rho\left[c(\partial X^\rho+2i\,p_{2,\sigma}\psi^\rho\psi^\sigma)-\gamma\psi^\rho\right]e^{2ip_2\cdot X}(z_2)\,,\\
 V_{G}(z_3)&=c\,e^{-\tfrac{\varphi}{2}}S_\alpha S_{A}e^{iP\cdot X}(z_3)\,,\\
 V_{G}(z_4)&=-\frac{i}{16\pi\sqrt{2}}\epsilon^{AB}(\sigma^{\lambda\kappa})^{\alpha\beta}F_{\lambda\kappa}e^{-\tfrac{\varphi}{2}}S_\beta S_{B}e^{iP\cdot X}(z_4)\,,\\
 V_{\bar F}(z_5)&=c\,e^{-\tfrac{\varphi}{2}}S_\gamma S_{\hat A}e^{i\bar P\cdot X}(z_5)\,,\\
 V_{\bar F}(z_6)&=\frac{i}{16\pi\sqrt{2}}\epsilon^{\hat A\hat B}(\sigma^{\omega\xi})^{\gamma\delta}\bar F_{\omega\xi}e^{-\tfrac{\varphi}{2}}S_\delta S_{\hat B}e^{i\bar P\cdot X}(z_6)\,.
\end{align}
Here, I have only written the R-R parts of the closed string vertices whose contributions are considered first. Furthermore, recall that momentum conservation along the Neumann directions implies that $p_1+p_2+P+\bar P=0$. Even though I compute a coupling at a fixed momentum order, I keep the momenta generic in order to consistently regularise the worldsheet integrals. Equivalently, the contribution of interest arises as a contact term of the form $(p_i\cdot p_j)/(p_i\cdot p_j)$ such that it is crucial to send the momenta to zero only at the end.

There are, in principle, several terms stemming from whether one takes the bosonic or fermionic parts of the gauge fields. First of all, one can show that the terms needed to reinforce BRST invariance of the gauge fields vertex operators do not lead to any non-trivial contributions. There are thus four separate CFT correlators all of which are multiplied by
\begin{align}
 A_0=\frac{1}{2^9\pi^2}\epsilon^{AB}\epsilon^{\hat A\hat B}A_{\mu}A_{\rho}(\sigma^{\lambda\kappa})^{\alpha\beta}F_{\lambda\kappa}(\sigma^{\omega\xi})^{\gamma\delta}\bar F_{\omega\xi}\frac{z_{12}z_{14}z_{24}}{(z_{34}z_{35}z_{36}z_{45}z_{46}z_{56})^{\tfrac{1}{4}}}\prod_{i,j=1\atop i<j}^{6}z_{ij}^{4p_i\cdot p_j}\,.
\end{align}
Here, I have defined $p_3=p_4=P/2$ and $p_5=p_6=\bar P/2$. If one takes both gauge fields to give their bosonic term, one obtains the term
\begin{align}
 A_1&=\left\langle\partial X^\mu(z_2)\partial X^\rho(z_2)\right\rangle\left\langle S_\alpha S_A(z_3)S_\beta S_B(z_4)S_\gamma S_{\hat A}(z_5)S_\delta S_{\hat B}(z_6)\right\rangle\nonumber\\
    &=-\frac{\epsilon_{AB}\epsilon_{\hat A\hat B}\epsilon_{\alpha\delta}\epsilon_{\beta\gamma}z_{36}z_{45}}{(z_{34}z_{35}z_{36}z_{45}z_{46}z_{56})^{\tfrac{3}{4}}}\left(2\frac{p_2^\mu}{z_{12}}+\frac{P^\mu}{z_{13}}+\frac{P^\mu}{z_{14}}+\frac{\bar P^\mu}{z_{15}}+\frac{\bar P^\mu}{z_{16}}\right)\left(2\frac{p_1^\rho}{z_{12}}-\frac{P^\rho}{z_{23}}-\frac{P^\rho}{z_{24}}-\frac{\bar P^\rho}{z_{25}}-\frac{\bar P^\rho}{z_{26}}\right)\,.
\end{align}
Notice that I have only kept the terms that would be non vanishing once I take into account the polarisation vectors in $A_0$. However, in the gauge chosen for the worldsheet positions, this term is zero by the transversality condition $p_1^\mu A_{\mu}=0$. The second term obtained when only the gauge field at $z_2$ gives its bosonic piece is
\begin{align}
 A_2&=2ip_{1,\nu}\left\langle\partial X^\rho(z_2)\right\rangle\left\langle\psi^\mu\psi^\nu(z_1)S_\alpha S_A(z_3)S_\beta S_B(z_4)S_\gamma S_{\hat A}(z_5)S_\delta S_{\hat B}(z_6)\right\rangle\nonumber\\
    &=\frac{-p_{1,\nu}\epsilon_{AB}\epsilon_{\hat A\hat B}z_{34}z_{56}}{z_{13}z_{14}z_{15}z_{16}(z_{34}z_{35}z_{36}z_{45}z_{46}z_{56})^{\tfrac{3}{4}}}\left[\epsilon_{\beta\gamma}(\sigma^{\mu\nu})_{\alpha\delta}z_{14}z_{15}z_{36}+\epsilon_{\alpha\delta}(\sigma^{\mu\nu})_{\beta\gamma}z_{13}z_{16}z_{45}\right]\nonumber\\
&\times\left(2\frac{p_1^\rho}{z_{12}}-\frac{P^\rho}{z_{23}}-\frac{P^\rho}{z_{24}}-\frac{\bar P^\rho}{z_{25}}-\frac{\bar P^\rho}{z_{26}}\right)
\end{align}
Similarly to $A_1$, when the gauge field at $z_1$ only gives its bosonic piece then the correlator vanishes. Finally, when all vertices give their fermionic terms, one finds
\begin{align}
 A_{4}=&-4p_{1,\nu}p_{2,\sigma}\left\langle\psi^{\mu}\psi^{\nu}(z_1)\psi^{\rho}\psi^{\sigma}(z_2)S_\alpha S_A(z_3)S_\beta S_B(z_4)S_\gamma S_{\hat A}(z_5)S_\delta S_{\hat B}(z_6)\right\rangle\nonumber\\
      =&\frac{2p_{1,\nu}p_{2,\sigma}\epsilon_{AB}\epsilon_{\hat A\hat B}}{z_{12}z_{13}z_{14}z_{15}z_{16}z_{23}z_{24}z_{25}z_{26}(z_{34}z_{35}z_{36}z_{45}z_{46}z_{56})^{\tfrac{3}{4}}}\Biggl[\delta^{\mu\rho}\epsilon_{\beta\gamma}\sigsig{\nu}{\sigma}{\alpha\delta}z_{13}z_{15}^2z_{24}^2z_{26}z_{34}z_{36}z_{56}\nonumber\\
  &+\delta^{\mu\rho}\epsilon_{\alpha\delta}\sigsig{\nu}{\sigma}{\gamma\beta}z_{13}^2z_{15}z_{24}z_{26}^2z_{34}z_{45}z_{56}+\delta^{\nu\sigma}\epsilon_{\beta\gamma}\sigsig{\mu}{\rho}{\alpha\delta}z_{13}z_{15}^2z_{24}^2z_{26}z_{34}z_{36}z_{56}\nonumber\\
  &+\delta^{\nu\sigma}\epsilon_{\alpha\delta}\sigsig{\mu}{\rho}{\gamma\beta}z_{13}^2z_{15}z_{24}z_{26}^2z_{34}z_{45}z_{56}-\delta^{\mu\sigma}\epsilon_{\beta\gamma}\sigsig{\nu}{\rho}{\alpha\delta}z_{13}z_{15}^2z_{24}^2z_{26}z_{34}z_{36}z_{56}\nonumber\\
  &-\delta^{\mu\sigma}\epsilon_{\alpha\delta}\sigsig{\nu}{\rho}{\gamma\beta}z_{13}^2z_{15}z_{24}z_{26}^2z_{34}z_{45}z_{56}-\delta^{\nu\rho}\epsilon_{\beta\gamma}\sigsig{\mu}{\sigma}{\alpha\delta}z_{13}z_{15}^2z_{24}^2z_{26}z_{34}z_{36}z_{56}\nonumber\\
  &-\delta^{\nu\rho}\epsilon_{\alpha\delta}\sigsig{\mu}{\sigma}{\gamma\beta}z_{13}^2z_{15}z_{24}z_{26}^2z_{34}z_{45}z_{56}\nonumber\\
  &-\frac{1}{2}\ssss{\mu}{\nu}{\rho}{\sigma}{\alpha\delta}\epsilon_{\beta\gamma}z_{12}z_{14}z_{15}z_{24}z_{25}z_{34}z_{36}^2z_{56}+\frac{1}{2}\ssss{\mu}{\nu}{\rho}{\sigma}{\gamma\beta}\epsilon_{\alpha\delta}z_{12}z_{13}z_{16}z_{23}z_{26}z_{34}z_{45}^2z_{56}\nonumber\\
  &+\frac{1}{2}z_{12}\left(\sigsig{\mu}{\nu}{\alpha\delta}\sigsig{\rho}{\sigma}{\gamma\beta}z_{12}z_{15}z_{26}z_{34}^2z_{36}z_{45}z_{56}+\sigsig{\mu}{\nu}{\gamma\beta}\sigsig{\rho}{\sigma}{\alpha\delta}z_{12}z_{13}z_{24}z_{34}z_{36}z_{45}z_{56}^2\right)\nonumber\\
  &-\frac{1}{2}z_{12}z_{13}z_{15}z_{24}z_{26}z_{34}z_{36}z_{45}z_{56}\left(\sigsig{\nu}{\rho}{\alpha\delta}\sigsig{\mu}{\sigma}{\gamma\beta}+\sigsig{\nu}{\rho}{\gamma\beta}\sigsig{\mu}{\sigma}{\alpha\delta}\right)\Biggr]\,.
\end{align}
Here, I have used the explicit results of \cite{Haertl:2009yf} for higher point fermionic correlation function on the disc. This can be further simplified to
\begin{align}
 A_{4}&=\frac{2p_{1,\nu}p_{2,\sigma}\epsilon_{AB}\epsilon_{\hat A\hat B}}{z_{12}z_{13}z_{14}z_{15}z_{16}z_{23}z_{24}z_{25}z_{26}(z_{34}z_{35}z_{36}z_{45}z_{46}z_{56})^{\tfrac{3}{4}}}\Biggl\{\nonumber\\
  &+z_{13}z_{15}^2z_{24}^2z_{26}z_{34}z_{36}z_{56}\left(\delta^{\mu\rho}\epsilon_{\beta\gamma}\sigsig{\nu}{\sigma}{\alpha\delta}+\delta^{\nu\sigma}\epsilon_{\beta\gamma}\sigsig{\mu}{\rho}{\alpha\delta}-\delta^{\mu\sigma}\epsilon_{\beta\gamma}\sigsig{\nu}{\rho}{\alpha\delta}-\delta^{\nu\rho}\epsilon_{\beta\gamma}\sigsig{\mu}{\sigma}{\alpha\delta}\right)\nonumber\\
  &+z_{13}^2z_{15}z_{24}z_{26}^2z_{34}z_{45}z_{56}\left(\delta^{\mu\rho}\epsilon_{\alpha\delta}\sigsig{\nu}{\sigma}{\gamma\beta}+\delta^{\nu\sigma}\epsilon_{\alpha\delta}\sigsig{\mu}{\rho}{\gamma\beta}-\delta^{\mu\sigma}\epsilon_{\alpha\delta}\sigsig{\nu}{\rho}{\gamma\beta}-\delta^{\nu\rho}\epsilon_{\alpha\delta}\sigsig{\mu}{\sigma}{\gamma\beta}\right)\nonumber\\
  &-\frac{1}{2}\ssss{\mu}{\nu}{\rho}{\sigma}{\alpha\delta}\epsilon_{\beta\gamma}z_{12}z_{14}z_{15}z_{24}z_{25}z_{34}z_{36}^2z_{56}+\frac{1}{2}\ssss{\mu}{\nu}{\rho}{\sigma}{\gamma\beta}\epsilon_{\alpha\delta}z_{12}z_{13}z_{16}z_{23}z_{26}z_{34}z_{45}^2z_{56}\nonumber\\
  &+\frac{1}{2}z_{12}z_{34}z_{36}z_{45}z_{56}\biggl[z_{12}z_{15}z_{26}z_{34}\sigsig{\mu}{\nu}{\alpha\delta}\sigsig{\rho}{\sigma}{\gamma\beta}+z_{12}z_{13}z_{24}z_{56}\sigsig{\mu}{\nu}{\gamma\beta}\sigsig{\rho}{\sigma}{\alpha\delta}\nonumber\\
  &-z_{13}z_{15}z_{24}z_{26}\left(\sigsig{\nu}{\rho}{\alpha\delta}\sigsig{\mu}{\sigma}{\gamma\beta}+\sigsig{\nu}{\rho}{\gamma\beta}\sigsig{\mu}{\sigma}{\alpha\delta}\right)\biggr]\Biggr\}\,.
\end{align}
The full correlator is thus $A_0(A_2+A_4)$ which I now integrate over the worldsheet positions $x$, $z_5$ and $z_6$. For generic space-time momenta, this is a well defined integral over $\mb R\times\mb C$. Instead of performing directly the calculation using the analytic structure of the integrand, I use the beautiful results of \cite{Stieberger:2009hq} in which this type of complex integrals is mapped to real integrals appearing in colour ordered amplitudes in gauge theory. More precisely, it was proven that
\begin{align}
 \int_{\mb R\times\mb C}\textrm{d}x\,&\textrm{d}^{2}z(z-\bar z)^\kappa x^{\alpha}(1+ix)^{u+n_0}(1-ix)^{u+n_1}(1-z)^{t+n_2}(1-\bar z)^{t+n_3}(z-ix)^{\tfrac{s}{2}+n_4}(\bar z+ix)^{\tfrac{s}{2}+n_5}\nonumber\\
 &\times(z+ix)^{\tfrac{s}{2}+n_6}(\bar z-ix)^{\tfrac{s}{2}+n_7}\nonumber\\
 =2\sin&(\frac{\pi s}{2})\sin(\pi s)A(1,6,3,5,4,2)-2\sin(\frac{\pi s}{2})\sin(\pi t)A(1,3,5,4,2,6)\,,
\end{align}
with
\begin{align}
 A(1,6,3,5,4,2)=&-\frac{1}{2}(-)^{n_4+n_5+n_7}\int_{0}^{1}\textrm{d}x\,x^{u-2-\kappa-n_3-n_5-n_7}(1-x)^{\tfrac{s}{2}+n_5}\int_{0}^{1}\textrm{d}y\,y^{-\tfrac{s}{2}+n_1+n_2+n_4+1}\nonumber\\
 &\times(1-y)^{\tfrac{s}{2}+n_6}(1-xy)^\kappa\int_{0}^{1}\textrm{d}z\,z^{u+n_1}(1-z)^{\tfrac{s}{2}+n_4}(1-yz)^{\alpha}(1-x y z)^{\tfrac{s}{2}+n_7}\,,
\end{align}
and
\begin{align}
 A(1,3,5,4,2,6)=&\frac{1}{2}(-)^{\kappa+n_3+n_4}\int_{0}^{1}\textrm{d}x\,x^{\tfrac{s}{2}+n_6}(1-x)^{\tfrac{s}{2}+n_4}\int_{0}^{1}\textrm{d}y\,y^{s+\alpha+n_4+n_6+1}(1-y)^{u+n_1}(1-xy)^{t+n_1}\nonumber\\
 &\times\int_{0}^{1}\textrm{d}z\,z^{u-\kappa-n_3-n_5-n_7-2}(1-z)^{t+n_3}(1-yz)^{\tfrac{s}{2}+n_6}\,.
\end{align}

These integrals are generically quite involved. However, recall that in the present case, I am only interested in the leading contribution in the space-time momenta. The latter stems from contact terms arising from the worldsheet integrals, \emph{i.e.} poles in $s$, $t$ or $u$ where
\begin{align}
s&=2p_1\cdot p_2=2P\cdot\bar P\,,\nonumber\\
s&=2p_1\cdot P=2p_2\cdot\bar P\,,\nonumber\\
s&=2p_1\cdot\bar P=2p_2\cdot P\,.
\end{align}
Once the leading contributions from the worldsheet integrals is extracted, one evaluates the total amplitude $A_0(A_2+A_4)$ in the limit of vanishing momenta. The identities for the traces of Pauli matrices derived in Appendix \ref{appendix:spinors} turn out to be very useful. The result takes the following simple form:
\begin{align}
  \left\langle\left\langle V_{A}(z_1)V_{A}(z_2)V_G^{(-\tfrac{1}{2},-\tfrac{1}{2})}(z_3,z_4)V_{S'}^{(-\tfrac{1}{2},-\tfrac{1}{2})}(z_5,z_6)\right\rangle\right\rangle=A_{\mu}A^\mu F_{G,\rho\sigma}F_{S'}^{\rho\sigma}\,.
\end{align}
One can perform the same analysis for all other possible terms, for instance when including the NS-NS parts of the vertex operators, and I find
\begin{align}\label{QuadYM}
  \left\langle V_{A}V_{A}V_GV_{S'}\right\rangle=\epsilon\,\bar\epsilon\,A_{\mu}A^\mu\,.
\end{align}

Before concluding this session, one should check that the full deformation does not completely break supersymmetry, even in the presence of $\bar\Omega$. More precisely, this boils down to showing that the scalar component of the gauge multiplet remains massless. In order to achieve that, one should calculate all possible couplings between two scalars $\phi,\bar\phi$ and the $\Omega,\bar\Omega$ deformation in the field theory limit. Instead of performing the disc amplitude calculations as before, I simply note that this precise analysis has already been done in \cite{Micha} where it was shown that the mass deformation realised at the level of the worldsheet does not lead to a mass term for the Higgs scalar.  The situation here is exactly the same by a mere exchange of the space-time $\mb C^2$ and the $\mb C^2$ of the internal space used in \cite{Micha}. It is important to note that the role of the quadratic deformation is crucial to ensure this property. Namely, as extensively studied in \cite{BarDec}, the full description of the $\Omega$-deformation requires the introduction of a background for the scalar field $S'$ at second order in momenta and proportional to $\epsilon\bar\epsilon$. As reviewed in Section \ref{Het}, in heterotic string theory, this additional ingredient implies an exact decoupling of $\bar\Omega$ from the topological amplitudes $F_g$ as a cancellation between bosonic and fermionic degrees of freedom. In the present case, it ensures that supersymmetry is not broken while it does not change the form of the remaining couplings.

To summarise our findings, I write down the full $\Omega$-deformed Yang-Mills action derived from the realisation of the deformation as a constant background of particular string states. The action is
\begin{align}
 \mc L_{\textrm{SYM}}=\frac{1}{g_{\textrm{YM}}^2}\textrm{Tr}&\Biggl\{\frac{1}{2}F^2-2\bar\Lambda_{\dot\alpha A}\left(\bar{\slashed{D}}^{\dot\alpha\beta}{\Lambda_{\beta}}^A+i\sqrt{2}\left[\phi,\bar\Lambda^{\dot\alpha A}\right]\right)+2i\sqrt{2}{\Lambda^{\alpha}}_{A}\left(\epsilon_{\alpha\beta}[\bar\phi,\Lambda_{\beta A}]+\tfrac{\bar\epsilon}{2}{\left(\sigma^3\right)_{\alpha}}^{\beta}{\Lambda_\beta}^{A}\right)\nonumber\\
&+\left(D_{\mu}\phi-\epsilon\,A^\nu\eta^3_{\mu\nu}\right)\left(D^{\mu}\bar\phi-\bar\epsilon\,A_\rho\eta^{3,\mu\rho}\right)\Biggr\}+\mc O(\alpha')\,,\label{SYMaction}
\end{align}
which shows that the realisation through the graviphoton and the $S'$-vector is consistent. In the following section, I prove that this statement holds non-perturbatively.

\subsection{Deformed ADHM action}

\begin{figure}[h!t]
\begin{center}
\includegraphics[width=0.8\textwidth]{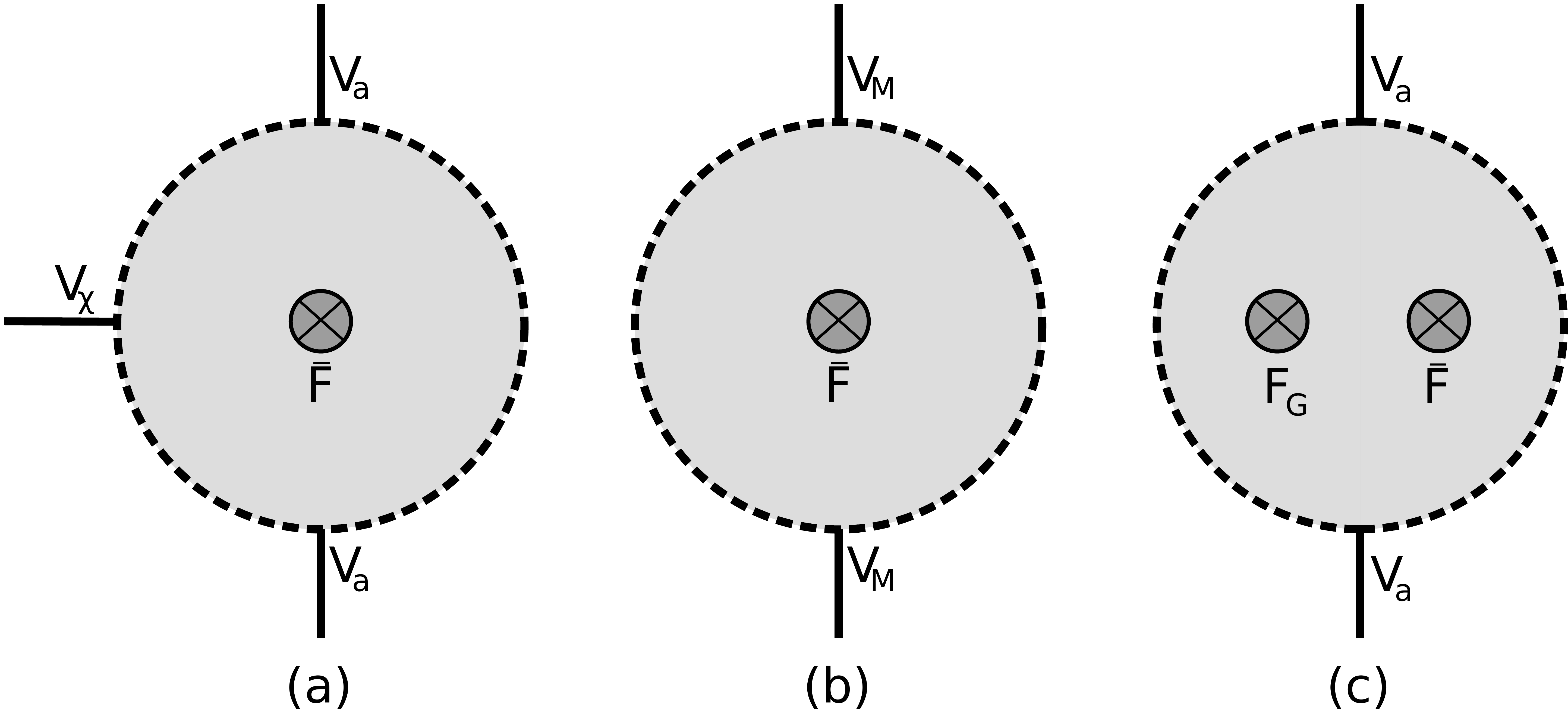}
\end{center}
\caption{Disc diagrams in the 5-5 sector. Diagram (a) involves three bosonic boundary insertions and one bulk insertion of $\bar F$, diagram (b) two fermionic insertions and one $\bar F$, whereas diagram (c) has two bosonic boundary insertions and two bulk insertions of a graviphoton and an $\bar F$.} 
\label{fig:Instdiags}
\end{figure}

I now focus on the deformation of the ADHM action stemming from the closed string vertices. As I now show, this confirms the choice of the vertex made previously and leads to the explicit expression of the $\bar\Omega$-deformed instanton effective action as a check.
As in the Yang-Mills case, there is a limited number of couplings surviving the field theory limit and self-duality projections. Some of these are summarised in Figure \ref{fig:Instdiags}.

In principle, there could as well be couplings with mixed D5/D9 boundary conditions. However, due to the self-duality constraint, such couplings are zero. In the case of a general $\Omega$-background, this is of course not any longer true \cite{Antoniadis:2013mna}.

I first consider the coupling involving fermionic ADHM moduli. That is, for the NS-NS part,
\begin{align}
 \left\langle\left\langle c V_{M}(z_1)c V_{M}(z_2)c V^{(-1,0)}_{\bar F}(z_3,z_4)\right\rangle\right\rangle&=\frac{1}{8\pi\,\sqrt{2}}M^{\alpha A}M^{\beta B}\bar F_{\mu\nu}\Big\langle\Big\langle c(z_1)S_{\alpha}(z_1)S_{A}(z_1)\,e^{-\tfrac{1}{2}\varphi(z_1)}\nonumber\\
&\times c(z_2)S_{\beta}(z_2)S_{B}(z_2)\,e^{-\tfrac{1}{2}\varphi(z_2)}\nonumber\\
&\times c(z_3)e^{-\varphi(z_3)}\bar\Psi(z_3)\psi^{\mu}\psi^{\nu}(z_4)\Big\rangle\Big\rangle+(\textrm{Left}\leftrightarrow\textrm{Right})\,.
\end{align}
Notice that, apart from the fact that space-time momenta are forbidden for the ADHM moduli due to the Dirichlet boundary conditions, the calculation goes along the same lines as \eqref{LambdaLambdaFsetup}. In addition, the $g_6^2$ factor from the vertices \eqref{MlambdaVop} cancels with the normalisation of the D5 amplitude. Therefore, I can immediately state the result. Namely, for the $S$ one has
\begin{equation}
 \Big\langle V_{M}V_{M}V_S\Big\rangle=0
\end{equation}
while
\begin{equation}
 \Big\langle V_{M}V_{M}V_{S'}\Big\rangle=-i\sqrt{2}\,\bar\epsilon M^{\alpha A}{M^\beta}_A(\sigma^3)_{\alpha\beta}\,.
\end{equation}

I now turn to the bosonic term. First consider the coupling corresponding to Diagram (a) in Fig. \ref{fig:Instdiags}. Once again, for simplicity, I choose the zero-picture vertices to be of dimension one (such that their positions are integrated),  and all the $(-1)$-picture vertices to be of dimension zero (such that their positions remain unintegrated) so that one must insert one PCO. As I show below, the amplitude takes the form of contact terms in the momenta of the form $p_i.p_j/p_i.p_j$. These contact terms survive in the limit $p_i\to 0$. To be able to compute them in a well-defined manner, the momenta $p_i$ must be kept generic as they also act as regulators of the worldsheet integrals. The limit of vanishing momenta is only taken at the end of the calculation. Notice, however, that due to the nature of our vertex insertions, none of the ADHM moduli can carry momenta along $X^\mu$ since the four-dimensional space-time corresponds to directions with Dirichlet boundary conditions for the D5-instantons. Similarly, the $S'$-vector insertions cannot carry momenta along $T^2$ because of BRST invariance. As a way out, I turn on complex momenta along the $K3$ directions for all the vertices to make all integrals well-defined. Technically, this means that I first replace $K3$ by $\mathbb{R}^4$ and compute the amplitude on the D-instanton world-volume $T^2 \times \mathbb{R}^4$. Since the fields appearing in this amplitude survive the orbifold projection\footnote{More generally on a smooth $K3$ manifold they give rise to zero-modes.}, the coupling of interest also exists in the case where $\mathbb{R}^4$ is replaced by $K3$.

As the role of space-time and internal momenta is crucial, and for the sake of clarity, I write the relevant vertex operators below.
\begin{eqnarray}
V_{a}(z_1)&=& g_6\,a_{\mu}(\partial X^\mu-2 i  p_1\cdot\chi \,\psi^{\mu})e^{2 i p_1\cdot Y}(z_1)\,,\label{Va}\\
V_{a}(z_2)&=& g_6\,a_{\nu}\, c e^{-\varphi}\psi^{\nu}e^{2 i p_2\cdot Y}(z_2)\,,\\
V_{\chi}(z_3)&=&\frac{\chi}{\sqrt{2}}\, (\partial Z-2 i p_3\cdot\chi \,\Psi)e^{2 i p_3\cdot Y}(z_3)\label{Vchi}\,,\\
V_{\bar F}(z)&=&  c e^{-\varphi}\bar{\Psi}e^{i(P_\mu X^\mu+ P\cdot Y)}(z)\,,\\
V_{\bar F}(\bar{z})&=&-\frac{i\epsilon_{\lambda}}{8\pi\sqrt{2}}\,  c e^{-\varphi}\psi^{\lambda}e^{i(-P_\mu X^\mu+ P\cdot Y)}(\bar{z})\,,
\end{eqnarray}
with $\bar F_{\mu\nu}\equiv \epsilon_{[\mu}P_{\nu]}$, and the only relevant terms in $V_{\text{PCO}}$ are
\begin{equation}
e^{\varphi}\, T_F(y)= ie^{\varphi}(\psi^\mu \partial X_\mu+\Psi \partial \bar{Z}+\bar{\Psi} \partial{Z}+\chi^i \partial Y^i)(y)\,.
\end{equation}
Here, $Y^i\in \{X^6,X^7,X^8,X^9\}$ are coordinates of the internal $\mathbb{R}^4$. The momenta $p_i$ are along these directions, while the momentum of $V_{F^{\bar{S}'}}$ is written as $(P_\mu, P)$, where $P_\mu$ is the space-time part and $P$ is along the $Y^i$ directions. Note that after using the doubling trick, the Neumann directions $( Z, \bar{Z},Y^i)$ are mapped onto themselves, whereas the Dirichlet ones pick an additional minus sign $X^\mu \rightarrow -X^{\mu}$. This is consistent with the fact that the momenta along Neumann directions are conserved. On the other hand, integrating over the zero modes of the Dirichlet directions $X^\mu$ does not give rise to any conservation law for the momentum $P_\mu$.

The three open string vertices contain Chan-Paton labels that should be suitably ordered. For instance, in order to obtain the term $\textrm{Tr}(a_{\mu}\,a_{\nu}\,\chi)$, the range of integration is
\begin{equation}\label{IntRange}
\begin{cases}
 \textrm{for } z_1>z_2\,, & z_3\in ]z_2,z_1[\,,\\
 \textrm{for } z_2>z_1\,, & z_3\in]-\infty,z_1[\cup]z_2,\infty[\,.
\end{cases}
\end{equation}
For the other nonequivalent ordering $\textrm{Tr}(a_{\mu}\,\chi\,a_{\nu})$, the range of integration in $z_3$ is opposite. It is easy to show that the sum of these two orderings gives $\textrm{Tr}(a_{\mu}[\chi, a_{\nu}])$.

For definiteness, I first focus on the term $\textrm{Tr}(a_{\mu}\,a_{\nu}\,\chi)$. The contractions of the ghosts, superghosts and the exponentials in momenta yield
\begin{eqnarray}
A_0&=&\left\{-\frac{ig_6^2}{16\pi}\,\textrm{Tr}(a_{\mu}\,a_{\nu}\,\chi)\epsilon_{\lambda}\right\}|y-z|^2(y-z_2)\nonumber\\
&\times&\prod_{1\leq i<j\leq3}(z_{ij})^{4 p_i\cdot p_j}\prod_{k=1}^{3}|z_k-z|^{4 p_k\cdot P}(z-\bar{z})^{- P_\mu P^\mu+ P_i P_i}\,.
\end{eqnarray}
This factor multiplies each of the remaining contractions. Now consider the contribution of $\partial{Z}(z_3)$ to the amplitude. This can only contract with $\partial \bar{Z}(y)$ in $V_{\textrm{PCO}}$ while $\Psi(y)$ contracts with $\bar\Psi(z)$. Hence, $\psi^{\lambda}(\bar{z})$ must contract with $\psi^\nu(z_2)$ and from $z_1$ only $\partial X^{\mu}(z_1)$ can contribute. The result is
\begin{equation}
A_1= \frac{i \delta^{\nu \lambda} P^\mu (z-\bar{z})}{(y-z_3)^2 (y-z)(z_2-\bar{z})|z_1-z|^2}\,.
\end{equation}
Next consider the contribution of the second term in \eqref{Vchi} where there are several contributions. First, if $p_3\cdot\chi(z_3)$ contracts with $p_1\cdot\chi(z_1)$, then $\psi^\mu(z_1)$, $\psi^{\nu}(z_2)$, $\psi^{\lambda}(\bar{z})$ and a space-time fermion $\psi^{\sigma}(y)$ from the PCO must contract, leaving $\partial X^\sigma(y)$ to contract with the momentum parts of the operators at $z$ and $\bar{z}$. This gives a term proportional to $P_\sigma$. Secondly, notice that the term arising from contracting $\psi^\mu$ with $\psi^{\nu}$ is killed by the transversality condition. The total result is
\begin{equation}
A_2= \frac{4 i p_1\cdot p_3\  (z-\bar{z})}{(z_3-z)z_{31}|y-z|^2}\left[\frac{\delta^{\nu \lambda} P^\mu}{(z_2-\bar{z})(y-x_1)}-\frac{\delta^{\mu \lambda} P^\nu}{(z_1-\bar{z})(y-z_2)}\right]\,.
\end{equation}
On the other hand, if the term $p_3\cdot\chi(z_3)$ contracts with $\chi(y)$ in $V_{\textrm{PCO}}$, $\partial Y(y)$ must contract with momentum dependent parts of the vertices. Thus, $\psi^{\lambda}(\bar{z})$ contracts with $\psi^\nu(z_2)$ and only $\partial X^\mu$ at $z_1$ contributes so that one obtains
\begin{equation}
A_3=\frac{4 i\delta^{\nu \lambda} P^\mu (z-\bar{z})}{(z_3-z)(y-z_3)(z_2-\bar{z})|z_1-z|^2}\left[\frac{p_3\cdot p_1}{y-z_1}+\frac{p_3\cdot p_2}{y-z_2}+\frac{p_3\cdot P}{2(y-z)}+
\frac{p_3\cdot P}{2(y-\bar{z})}\right]\,.
\end{equation}
The total correlation function is thus given by 
\begin{equation}
 \Big\langle V_a(z_1)V_a(z_2)V_\chi(z_3)V_{\bar F}^{\textrm{NS-NS}}\Big\rangle=A_0(A_1+A_2+A_3)\,,
\end{equation}
which I now integrate over $z_1$ and $z_3$. Note that all the terms in $A_1$, $A_2$ and $A_3$ come with a single power of space-time momentum $P^\mu$ which is exactly what is required to obtain a coupling to the field strength of the closed string state. However, both $A_2$ and $A_3$ are quadratic in the momenta along the $Y^i$ directions. Hence, they can only contribute to the amplitude in the zero-momentum limit if the integration over $z_1$ and $z_3$ gives a pole $1/(p_a\cdotp p_b)$. Clearly, $A_0\cdotp A_3$ cannot give such a pole\footnote{I am assuming a generic value of $y$ in the complex plane.}. On the other hand, the integral over $z_3$ for  $A_0\cdotp A_2$ gives a $1/(p_1\cdotp p_3)$ pole. Performing the $z_3$ integral in the regions \eqref{IntRange} yields the same result. Consequently, the $z_1$ integral over the entire real line reads
\begin{equation}
A_0 A_2 = \frac{g_6^2}{8\pi}\,\textrm{Tr}\left[a_{\mu}\,a_{\nu}\,\chi\right]\int_{-\infty}^{\infty} dz_1 \frac{(z-\bar{z})}{|z_1-z|^2}\epsilon_{\lambda}\left[\frac{(z_1-\bar{z})(y-z_2)}{(z_2-\bar{z})(y-z_1)}P^{\mu}\delta^{\nu\lambda}-P^{\nu}\delta^{\mu\lambda}\right]\,,
\label{A0A2}
\end{equation}
where I have set all the momenta along the $Y^i$ directions to zero given that there are no singularities in the remaining integral. Note that the integral $A_0\cdotp A_2$ is not gauge invariant. As for the $A_0 A_1$ term the $z_3$ and $z_1$ integrals contain no singularities so that the momenta along the $Y^i$ directions can be set to zero. The resulting $z_3$ integral for both regions \eqref{IntRange} gives the same result:
\begin{equation}
A_0 A_1 = -\frac{g_6^2}{8\pi}\,\textrm{Tr}\left[a_{\mu}\,a_{\nu}\,\chi\right]\int_{-\infty}^{\infty} dx_1 \frac{(z-\bar{z})}{|z_1-z|^2}\epsilon_{\lambda}\frac{(y-\bar{z})z_{12}}{(z_2-\bar{z})(y-z_1)}P^{\mu}\delta^{\nu\lambda}\,.\label{A0A1}
\end{equation}
Adding the two terms \eqref{A0A2} and \eqref{A0A1}, the total result becomes gauge invariant and the $z_1$ integration yields\footnote{I included an additional factor of 2 stemming from the left-right symmetrisation in the closed string vertex.}:
\begin{align}
 \Big\langle\Big\langle V_{a}\,V_{a}\,V_{\chi}\,V_{\bar F}^{\textrm{NS-NS}}\Big\rangle\Big\rangle&=-2i\,\textrm{Tr}\left[a_{\mu}\,a_{\nu}\,\chi\right]\epsilon_{\lambda}(P^{\mu}\delta^{\nu\lambda}-P^{\nu}\delta^{\mu\lambda})\,.\label{Phys4ptNS}
\end{align}
Finally, let us consider the R-R contributions. The vertex operators are the same as above, except for the closed string vertex:
\begin{align}
V^{(-\frac{1}{2},-\frac{1}{2})}_{\bar F}(z,\bar{z})=&\frac{i}{16\pi\sqrt{2}}\,\bar F^{\rho\lambda}\,c\,e^{-\frac{\varphi}{2}}\,S_{\alpha}\,S_{\hat A}\,e^{i(P\cdotp X+P\cdotp Y)}(z)\nonumber\\
&\times\epsilon^{\hat A\hat B}{({\sigma}_{\rho \lambda})^{\alpha}}_{\beta}\, 
c\,e^{-\frac{\varphi}{2}}\,S^{\beta}\,S_{\hat B}\,e^{i(-P\cdotp X+P\cdotp Y)}(\bar{z})\,.
\end{align}
Since the total superghost charge is $-2$, there is no need for a PCO. The total $T^2$ charge implies that only $p_3\cdot\chi\,\Psi(x_3)$ in \eqref{Vchi} contributes and, thus, only $p_1\cdot\chi\,\psi^{\mu}(x_1)$ in \eqref{Va} contributes. This term is proportional to $p_1\cdotp p_3$. As before, the integral over $z_3$ gives a pole $1/(p_1\cdotp p_3)$ in the channel where $z_3 \rightarrow z_1$. Performing the integrals over $z_1$ and $z_3$ gives the same result:
\begin{align}
 \Big\langle\Big\langle V_{a}\,V_{a}\,V_{\chi}\,V_{\bar F}^{\textrm{NS-NS}}\Big\rangle\Big\rangle&=\Big\langle\Big\langle V_{a}\,V_{a}\,V_{\chi}\,V_{\bar F}^{\textrm{R-R}}\Big\rangle\Big\rangle\,.
\end{align}
Finally, summing over the nonequivalent orderings of the open vertex operators gives
\begin{equation}\label{UnmixedPhys}
 \Big\langle\Big\langle V_{a}\,V_{a}\,V_{\chi}\,V_{\bar F}\Big\rangle\Big\rangle=-4i\,\textrm{Tr}\left[\chi,a_{\mu}\right]a_{\nu}\,F_{S'}^{\mu\nu}\,,
\end{equation}
while for $V_{S}$ it is zero as before!

To complete the analysis, I now calculate the coupling involving both $\Omega$ and $\bar\Omega$. This is the exact counterpart of \ref{QuadYM} for the instanton sector. Thus, the calculation goes along the same lines as the one leading to \ref{QuadYM}, with the subtle difference that, in the present case, momenta can only be turned on in the $K3$ directions. Consequently, I obtain a non-trivial quadratic coupling with $\Omega$ and $\bar\Omega$ making the instanton modulus $a$ massive, and the full deformation of the ADHM action including both self-dual $\Omega$ and $\bar\Omega$ is
\begin{align}
 S_{\textrm{ADHM}}=-\textrm{Tr}&\left\{\Big([\chi^{\dag},a_{\alpha\dot\beta}]+\bar\epsilon(\tau_3a)_{\alpha\dot\beta}\Big)\left([\chi,a^{\dot\beta\alpha}]+\epsilon(a\tau_3)^{\dot\beta\alpha}\right)+\frac{1}{2}M^{\alpha A}\Big(\left[\chi^\dag,M_{\alpha A}\right]+\bar\epsilon{(\tau_3)_\alpha}^{\beta}M_{\beta A}\Big)\right\}\,.
\end{align}
Clearly, I have only displayed the $\Omega,\bar\Omega$ dependent terms.


\section{\texorpdfstring{$\bar\Omega$ and topological amplitudes}{OmegaBar and topological amplitudes}}\label{Het}

\subsection{Effective action}

I now briefly review the results of \cite{BarDec} for completeness. There, based on the fact that parts of the topological amplitudes $F_g$ can be computed perturbatively in heterotic at one-loop using the couplings \cite{Antoniadis:1993ze}
\begin{align}
\mathcal{I}_g=&\int d^4x \int d^4\theta\, \mathcal{F}_g(X)\,(W_{\mu\nu}^{ij} W_{ij}^{\mu\nu})^g &&\text{for} && g\geq1\,,  \label{BpsStandard}
\end{align}
we proposed a deformation of the latter to capture also $\bar\Omega$ as a constant field strength for the vector partner of the K\"ahler modulus of $T^2$ denoted $F_T$\footnote{Recall that the dual heterotic theory is also compactified on $T^2\times K3$.}. Recall that $W_{\mu\nu}^{ij}$ is the supergravity multiplet carrying (anti-symmetrised) indices $i,j=1,2$ for the $SU(2)_R$ R-symmetry group. It contains the graviphoton field-strength $F^G$, the field strength tensor $B_{\mu\nu}^i$ of an $SU(2)$ doublet of gravitini and the Riemann tensor.

In order to include additional insertions of $F_T$, I deform the coupling $I_g$ as
\begin{align}
\mathcal{I}_{g,n}=&\int d^4x \int d^4\theta\, \mathcal{F}_{g,n}(X)\,W^{2g} K_T^{2n} \,,  \label{BpsNew}
\end{align}
where $K_T$ is the descendent superfield whose lowest component is $F_T$. Similarly to the previous type I analysis, it turns out that, in order to obtain a good description of $\bar\Omega$, one should also include an arbitrary number of the scalar field $T$ at two derivatives. At the level of the worldsheet sigma model, this translates into a quadratic deformation proportional to $\epsilon\bar\epsilon$ as before. 

\subsection{Amplitude calculation}

In order to calculate the coupling $F_{g,n}$, recall that the new couplings \eqref{BpsNew} are naturally calculated perturbatively at one-loop. For convenience, I choose particular kinematics in which the states of interest carry space-time momenta along $Z^1$ and $\bar Z^2$ only. The vertex operators for the graviphotons are
\begin{align}
 V_{G}(z_i)&=(\partial X-i\bar p_1\psi^2\Psi)\bar\partial Z^2\,e^{i\bar p_1 Z^1}\,,\nonumber\\
 V_{G}(w_i)&=(\partial X-ip_2\bar\psi^2\Psi)\bar\partial\bar Z^1\,e^{ip_2\bar Z^2}\,,
\end{align}
while for the $T$-vectors these are
\begin{align}
 V_{T}(s_i)&=(\partial Z^2-i\bar p_1\psi^1\psi^2)\bar\partial\bar X\,e^{i\bar p_1 Z^1}\,,\nonumber\\
 V_{T}(t_i)&=(\partial \bar Z^2-i p_2\bar\psi^2\bar\psi^1)\bar\partial\bar X\,e^{i p_2\bar Z^2}\,.
\end{align}                                                                                                                                                                                     
For simplicity, I choose the term in \eqref{BpsNew} in which there are $2g$ graviphotons and $2n+2$ field strengths $V_T$ so that the space-time zero-modes are soaked up by two vertices $V_T$. Hence, the amplitude to calculate is
\begin{equation}
 \left\langle V_T(x)V_T(y)\prod_{i=1}^{g} V_G(z_i)V_G(w_i)\prod_{j=1}^n V_{T}(s_i)V_{T}(t_i)\prod_{k=1}^m V_{\phi}(u_k)\right\rangle\,,
\end{equation}
which calculates the second derivative of $F_{g,n}$ with respect to $T$. Here, $V_{\phi}$ is the vertex operators of the scalar field $T$. As explained in \cite{BarDec}, after soaking up the space-time zero-modes, one finds that the coupling is integrable in the sense that one can pull out the two derivatives with respect to $T$. Furthermore,  by summing over $g$, $n$ and $m$, one can define a generating function in which all the amplitudes with arbitrary number of fields reduce to a Gaussian deformation of the worldsheet sigma-model! That is, the generating function
\begin{align}\label{GenSim}
 F(\epsilon,\bar\epsilon)=\Bigg\langle \textrm{exp}\Big[&-\epsilon\int\textrm{d}^2z\,\left(\partial XZ^1\bar\partial Z^2+\partial X\bar Z^2\bar\partial\bar Z^1\right)\nonumber\\
&-\bar\epsilon\int\textrm{d}^2z\,\left((Z^1\partial Z^2+\psi^1\psi^2)\bar\partial\bar X+(\bar Z^2\partial\bar Z^1+\bar\psi^1\bar\psi^2)\bar\partial\bar X\right)+\epsilon\bar\epsilon\int\textrm{d}^2z\,|Z^i|^2\partial X\bar\partial\bar X\Big]\Bigg\rangle\,.
\end{align}
calculates at once all the couplings $F_{g,n}$ which in turn can be recovered by picking a particular power of $\epsilon$ and $\bar\epsilon$. After a careful analysis of \eqref{GenSim}, one shows that the generating function is independent of $\bar\epsilon$. Indeed, it is given by
\begin{align}\label{FullAmplitudeHet}
 F\textrm(\epsilon)&=\int_{\mathcal{F}}\frac{d^2\tau}{\tau_2}\,\left(\frac{2\pi\epsilon\bar\eta^3}{\bar\theta_1(\pi\tilde\epsilon)}\right)^2\textrm{e}^{-\frac{\pi}{\tau_2}\tilde\epsilon^2}\frac{1}{\eta^4\bar\eta^{24}}\frac{1}{2}\sum_{h,g=0}^1 
G^\textrm{ferm}[^h_g](\check\epsilon_+)Z[^h_g]~\Gamma_{(2,2+8)}(T,U,Y)\,.
\end{align}
The details of the notation and the derivation can be found in \cite{BarDec}. In particular, expanding in the deformation parameters, one finds that
\begin{equation}
 \boxed{F_{g,n}=F_{g}}\,,
\end{equation}
with $F_g$ denoting here the heterotic one-loop topological amplitude.


\section{Conclusions}\label{Conclusion}

In the present paper, I have realised the $\bar\Omega$-deformation, the complex conjugate of the $\Omega$-deformation, in string theory in terms of a physical field in the string spectrum. In type I string theory compactified on $T^2\times K3$, that is the vector partner of the $S'$ scalar which describes the coupling of the D5-branes. By coupling this field to the graviphoton as well as to all the massless degrees of freedom of the D5-D9 system, I have derived the deformed Yang-Mills and ADHM effective actions. As already noticed in \cite{BarDec}, it is crucial to include a quadratic deformation which corresponds to giving a background to the $S'$ scalar at quadratic order in the momenta. This proves that the combination of graviphoton and $S'$-vector, together with the quadratic background, is a correct description of the full $\Omega$-background.

Furthermore, as already shown in \cite{BarDec}, one can go beyond the pure field theory analysis by calculating, in the dual heterotic theory, the topological amplitudes $F_g$ deformed by the $\bar\Omega$-deformation. Surprisingly, not only one recovers the perturbative part of the Nekrasov partition function, but also the full string result is independent of $\bar\Omega$. It would be interesting to understand this statement purely at the string level as a $Q$-exactness of some operator, as in the gauge theory.

From the conceptual point of view, the present analysis shows that the graviphoton differs from the $\Omega$-deformation. Indeed, it is clear that the quadratic deformation is essential to give a correct description of the full non-holomorphic $\Omega$-deformation. Yet, this additional piece corresponds to a different field in the string spectrum. However, $\Omega$ and the graviphoton agree at linear order and this is why, practically, one can neglect this subtlety.

As a natural application, it would be interesting to generalise my study beyond the topological limit. This is a priori tedious since, generically, the full $\Omega$-deformation in string theory breaks topological invariance. Yet, simplifications must occur once the complete $\bar\Omega$-deformation is included since the field theory limit is purely holomorphic. It is not clear, however, that the decoupling at the string level would still hold.


\section*{Acknowledgements}

I would like to thank O. Schlotterer for useful discussions. My work is supported by the Swiss National Science Foundation.

\appendix

\section{Notations and conventions}\label{appendix:spinors}
\subsection{Spinors}

I present some of our notations and conventions. $SO(4)$ spinor indices are raised and lowered using the standard epsilon tensors
\begin{align}
 &\epsilon^{12}=\epsilon_{12} =
1\,,\,\epsilon^{\dot{1}\dot{2}}=\epsilon_{\dot{1}\dot{2}} = -1\,,
\end{align}
such that
\begin{align}
&\psi^a = +\epsilon^{ab}\psi_b
\,,&&\psi_a=-\epsilon_{ab}\psi^b\,,&&\psi^{\dot a} = -\epsilon^{\dot a\dot
b}\psi_{\dot b} \,,&&\psi_{\dot a}=+\epsilon_{\dot a\dot b}\psi^{\dot b}\,.
\end{align}
In addition, $\sigma$-matrices for $SO(4)$, $(\sigma^k)_{a\dot a}$ and
$(\bar\sigma^k)^{\dot aa}$, are defined as
\begin{align}
	\sigma^\mu = (1\!\!1,-i\boldsymbol{\sigma})\,, && \bar\sigma^k =
(1\!\!1,+i\boldsymbol{\sigma})\,,
\end{align}
and are related by transposition. On the other hand, I denote $S_{\alpha}$ (resp. $S_{\dot\alpha}$) the self-dual (resp. anti-self-dual) spin fields of the space-time $SO(4)$. The spin fields for the internal manifold are, instead, $S_A$, $S^A$, $S_{\hat A}$, $S^{\hat A}$. Notice that covariant and contravariant indices $(A,\hat{A})$ of $SO(2)_{\pm}$ reflect charges $\pm 1/2$ with respect to $SO(2)$ according to the
decomposition $SO(6)\rightarrow SO(2)\times SO(4)$. Our conventions for the internal spin fields can be found in the table below.
\begin{align}\renewcommand{\arraystretch}{1.4}
	\begin{array}{c |c| c}
					\textrm{Spin field}  & SO(2) & SO(4) \\
\hline \hline
					S_A & + & (--),(++)\\ \hline
					S^A & - & (++),(--) \\ \hline
					S_{\hat{A}} & - & (-+),(+-) \\ \hline 
					S^{\hat{A}} & + & (+-),(-+) \\ \hline
	\end{array}\label{TabConventions}
\end{align}
In addition, the Lorentz generators $\sigma^{\mu\nu}, \bar\sigma^{\mu\nu}$ of $SO(4)_{ST}$ are
\begin{align}
{(\sigma_{\mu\nu})_\alpha}^\beta \equiv \frac{1}{2}{\bigr( \sigma_\mu\bar\sigma_\nu-\sigma_\nu\bar\sigma_\mu\bigr)_{\alpha}}^{\beta}\,,&&{(\bar\sigma_{\mu\nu})^{\dot\alpha}}_{\dot\beta} \equiv \frac{1}{2}{\bigr( \bar\sigma_\mu\sigma_\nu-\bar\sigma_\nu\sigma_\mu\bigr)^{\dot\alpha}}_{\dot\beta}\,.
\end{align}
They are symmetric in the spinor indices $(\sigma_{\mu\nu})_{\alpha\beta}=+(\sigma_{\mu\nu})_{\beta\alpha}$ and $(\bar\sigma_{\mu\nu})_{\dot\alpha\dot\beta}=+(\sigma_{\mu\nu})_{\dot\beta\dot\alpha}$. Also, they are (anti-)self-dual in the sense
\begin{align}
(\sigma^{\mu\nu})_{\alpha\beta} = +\frac{1}{2}\epsilon^{\mu\nu\rho\sigma} (\sigma_{\rho\sigma})_{\alpha\beta}\,, &&(\bar\sigma^{\mu\nu})_{\dot\alpha\dot\beta} = -\frac{1}{2}\epsilon^{\mu\nu\rho\sigma} (\bar\sigma_{\rho\sigma})_{\dot\alpha\dot\beta}\,.
\end{align}
Therefore, I can use them to define \emph{(anti-)self-dual} tensors. For example, for the self-dual $F_{\mu\nu}^{(+)}$ and anti-self-dual $F_{\mu\nu}^{(-)}$ part of the field strength tensor of a given gauge field, I write
\begin{align}
F^{(+)}_{\dot\alpha\dot\beta}\equiv (\bar\sigma^{\mu\nu})_{\dot\alpha\dot\beta}F^{(+)}_{\mu\nu}\,,&&F^{(-)}_{\alpha\beta}\equiv (\sigma^{\mu\nu})_{\alpha\beta} F^{(-)}_{\mu\nu}\,.\label{SelfDualDef}
\end{align}
Indeed, $F^{(\pm)}_{\mu\nu} = \mp \frac{1}{2} \epsilon_{\mu\nu\rho\sigma} (F^{(\pm)})^{\rho\sigma}$. Also, note the following useful identity:
\begin{align}
	&\sigma^{\mu\nu}\sigma^{\rho\sigma} = -(\delta^{\mu\rho}\delta^{\nu\sigma}-\delta^{\mu\sigma}\delta^{\nu\rho}+\epsilon^{\mu\nu\rho\sigma})\textrm{Id}_2+(\delta^{\sigma[\mu}\sigma^{\nu]\rho}-\delta^{\rho[\mu}\sigma^{\nu]\sigma})\,,
\end{align}
from which I can derive the following identities involving the traces of several Lorentz generators. For instance, for three generators, one has
\begin{align} \textrm{Tr}(\sigma^{\mu\nu}\sigma^{\rho\sigma}\sigma^{\lambda\delta})=2\delta^{\sigma[\mu}\left(\delta^{\nu]\delta}\delta^{\rho\lambda}-\delta^{\nu]\lambda}\delta^{\rho\delta}-\epsilon^{\nu]\rho\lambda\delta}\right)-2\delta^{\rho[\mu}\left(\delta^{\nu]\delta}\delta^{\sigma\lambda}-\delta^{\nu]\lambda}\delta^{\sigma\delta}-\epsilon^{\nu]\sigma\lambda\delta}\right)\,,
\end{align}
whereas for four generators I show that
\begin{align} \textrm{Tr}(\sigma^{\mu\nu}\sigma^{\rho\sigma}\sigma^{\lambda\delta}\sigma^{\kappa\tau})=&2(\delta^{\mu\rho}\delta^{\nu\sigma}-\delta^{\mu\sigma}\delta^{\nu\rho}+\epsilon^{\mu\nu\rho\sigma})(\delta^{\lambda\kappa}\delta^{\delta\tau}-\delta^{\lambda\tau}\delta^{\delta\kappa}+\epsilon^{\lambda\delta\kappa\tau})\nonumber\\
+&2\delta^{\sigma[\mu}\left(\delta^{\nu]\kappa}\delta^{\rho[\delta}-\delta^{\nu][\delta}\delta^{\rho\kappa}+\epsilon^{\nu]\rho\kappa[\delta}\right)\delta^{\lambda]\tau}
+2\delta^{\rho[\mu}\left(\delta^{\nu]\tau}\delta^{\sigma[\delta}-\delta^{\nu][\delta}\delta^{\sigma\tau}+\epsilon^{\nu]\sigma\tau[\delta}\right)\delta^{\lambda]\kappa}\nonumber\\
-&2\delta^{\rho[\mu}\left(\delta^{\nu]\kappa}\delta^{\sigma[\delta}-\delta^{\nu][\delta}\delta^{\sigma\kappa}+\epsilon^{\nu]\sigma\kappa[\delta}\right)\delta^{\lambda]\tau}
-2\delta^{\sigma[\mu}\left(\delta^{\nu]\tau}\delta^{\rho[\delta}-\delta^{\nu][\delta}\delta^{\rho\tau}+\epsilon^{\nu]\rho\tau[\delta}\right)\delta^{\lambda]\kappa}\,,
\end{align}

\subsection{Operator product expansions}\label{OPEs}

The operator product expansion algebra for the ten-dimensional fields can be decomposed according to the compactified theory. Indeed, the space-time current algebra is
\begin{alignat}{5}
&S^{\dot\alpha}(z)S_{\beta}(w)\,&\sim&\,\frac{1}{\sqrt{2}}\,{(\bar\sigma^\mu)^{
\dot\alpha}}_{\beta}\psi_\mu(w)\,,\,&S_{\alpha}(z)S^{\dot\beta}(w)\,&\sim\,\frac
{1}{\sqrt{2}}\,{(\sigma^\mu)_{\alpha}}^{\dot\beta}\psi_\mu(w)\,,\\
&S^{\dot\alpha}(z)S^{\dot\beta}(w)\,&\sim&\,-\frac{\epsilon^{\dot\alpha\dot\beta
}}{(z-w)^{1/2}}\,,\,&S_{\alpha}(z)S_{\beta}(w)\,&\sim\,\frac{\epsilon_{
\alpha\beta}}{(z-w)^{1/2}}\,,\\
&\psi^{\mu}(z)S^{\dot\alpha}(w)\,&\sim&\,\frac{1}{\sqrt{2}}\frac{
(\bar\sigma^\mu)^{\dot\alpha\beta}S_{\beta}(w)}{(z-w)^{1/2}}\,,\,&\psi^{\mu}
(z)S_{\alpha}(w)\,&\sim\,\frac{1}{\sqrt{2}}\frac{(\sigma^\mu)_{\alpha\dot\beta}
S^{\dot\beta}(w)}{(z-w)^{1/2}}\,,\\
&J^{\mu\nu}(z)S^{\dot\alpha}(w)\,&\sim&\,-\frac{1}{2}\frac{{(\bar\sigma^{\mu\nu}
)^{\dot\alpha}}_{\dot\beta}S^{\dot\beta}(w)}{z-w}\,,\,\,\,&J^{\mu\nu}(z)S_{
\alpha}(w)\,&\sim\,-\frac{1}{2}\frac{{(\sigma^{\mu\nu})_{\alpha}}^{\beta}S_{
\beta}(w)}{z-w}\,.
\end{alignat}
As for the internal one, it is given by
\begin{alignat}{5}
&S^{A}(z)S_{B}(w)\,&\sim&\,\frac{i{\delta^{A}}_{B}}{(z-w)^{3/4}}\,,\,&S_{A}(z)S^{B}(w)\,&\sim\,\frac{i{\delta_{A}}^{B}}{(z-w)^{3/4}}\\
&S^{A}(z)S^{B}(w)\,&\sim&\,\frac{i}{\sqrt{2}}\frac{(\Sigma^I)^{AB}\psi_{I}}{(z-w)^{1/4}}\,,\,&S_{A}(z)S_{B}(w)\,&\sim\,-\frac{i}{\sqrt{2}}\frac{(\Sigma^I)_{AB}\psi_{I}}{(z-w)^{1/4}}\,,\\
&\psi^{I}(z)S_{A}(w)\,&\sim&\,\frac{1}{\sqrt{2}}\frac{(\bar\Sigma^I)_{AB}S^{B}(w)}{(z-w)^{1/2}}\,,\,&\psi^{I}(z)S^{A}(w)\,&\sim\,-\frac{1}{\sqrt{2}}\frac{(\Sigma^I)^{AB}S_{B}(w)}{(z-w)^{1/2}}\,,\\
&J^{IJ}(z)S^{A}(w)\,&\sim&\,\frac{1}{2}\frac{{(\bar\Sigma^{IJ})^{A}}_{B}S^{B}(w)}{z-w}\,,\,\,\,&J^{IJ}(z)S_{A}(w)\,&\sim\,\frac{1}{2}\frac{{(\Sigma^{IJ})_{A}}^{B}S_{B}(w)}{z-w}\,.
\end{alignat}
Using the algebras above, one easily derives all the necessary correlation functions used throughout the manuscript. For instance, the correlation function an internal fermion with two spin fields is
\begin{equation}
 \left\<\Psi(z_1)S_{\hat A}(z_2)S_{\hat B}(z_3)\right\>=\frac{\epsilon_{\hat A\hat B}}{z_{12}^{\frac{1}{2}}z_{13}^{\frac{1}{2}}z_{23}^{\frac{1}{4}}}\,,
\end{equation}
whereas the correlator of two space-time fermions with two spin-fields is
\begin{equation}
 \left\<\psi^{\mu}(z_1)\psi^{\nu}(z_2)S_{\alpha}(z_3)S_{\beta}(z_4)\right\>=-\frac{1}{2}\frac{{\sigma^{\mu\nu}}_{\alpha\beta}z_{34}^{\frac{1}{2}}}{\left(z_{13}z_{14}z_{23}z_{24}\right)^{\frac{1}{2}}}+\frac{1}{2}\frac{\delta^{\mu\nu}\epsilon_{\alpha\beta}\left(z_{13}z_{24}+z_{14}z_{23}\right)}{z_{12}\left(z_{13}z_{14}z_{23}z_{24}z_{34}\right)^{\frac{1}{2}}}\,.
\end{equation}
Another useful result is the correlation function of four spin fields:
\begin{equation}
 \left\<S_{\alpha}(z_1)S_{\beta}(z_2)S_{\gamma}(z_3)S_{\delta}(z_4)\right\>=\frac{\epsilon_{\alpha\beta}\epsilon_{\gamma\delta}z_{14}z_{23}-\epsilon_{\alpha\delta}\epsilon_{\beta\gamma}z_{13}z_{24}}{\left(z_{12}z_{13}z_{14}z_{23}z_{24}z_{34}\right)^{\frac{1}{2}}}\,.
\end{equation}	
If one additionally inserts a space-time current, then
\begin{equation}
 \left\<\psi^{\mu}\psi^{\nu}(z_1)S_{\alpha}(z_2)S_{\beta}(z_3)S_{\gamma}(z_4)S_{\delta}(z_5)\right\>=\frac{1}{2}\frac{\epsilon_{\beta\gamma}(\sigma^{\mu\nu})_{\alpha\delta}z_{14}z_{15}z_{36}+\epsilon_{\alpha\delta}(\sigma^{\mu\nu})_{\beta\gamma}z_{13}z_{16}z_{45}}{z_{12}z_{13}z_{14}z_{15}\left(z_{23}z_{24}z_{25}z_{34}z_{35}z_{45}\right)^{\frac{1}{2}}}\,.
\end{equation}
Correlation functions for higher numbers of fermionic fields can be found in \cite{Haertl:2009yf}.

Finally, the two-point function for bosonic fields with Dirichlet boundary conditions is
\begin{align}
 \left\<\Pa Z_{\textrm{DD}}(z_1)\Pa \bar
Z_{\textrm{DD}}(z_2)\right\>=&\frac{1}{z_{12}^2}\,,\\
 \left\<\Pa Z_{\textrm{DD}}(z_1)\Bp \bar
Z_{\textrm{DD}}(z_2)\right\>=&-\frac{1}{z_{1\bar2}^2}\,,\\
 \left\<\Bp Z_{\textrm{DD}}(z_1)\Bp \bar
Z_{\textrm{DD}}(z_2)\right\>=&\frac{1}{z_{\bar1\bar2}^2}\,,
\end{align}
which can be found using
\begin{equation}
 \left\<Z_{\textrm{DD}}(z_1)\bar
Z_{\textrm{DD}}(z_2)\right\>=\log|z_{12}|^2-\log|z_{1\bar2}|^2\,.
\end{equation}
For the N-N directions, the same result holds with positive sign for all the two-point functions. For the fermions in the NS sector, it is the opposite:
\begin{align}
 \left\<\psi_{NN}(z_1)\psi_{NN}(z_2)\right\>=&\frac{1}{z_{12}}\,,&\,\left\<\psi_
{NN}(z_1)\psi_{NN}(\bar z_2)\right\>=&-\frac{1}{z_{1\bar2}}\,,\\
 \left\<\psi_{DD}(z_1)\psi_{DD}(z_2)\right\>=&\frac{1}{z_{12}}\,,&\,\left\<\psi_
{DD}(z_1)\psi_{DD}(\bar z_2)\right\>=&\frac{1}{z_{1\bar2}}\,.
\end{align}
\newpage

\bibliographystyle{utphys}
\bibliography{referencesv2}
\end{document}